\documentclass[showpacs, preprintnumbers, showkeys, nofootinbib]{revtex4}
\usepackage{amsmath, amsfonts, amssymb}
\usepackage{hyperref}

\newcommand{\N}{\mathcal{N}}
\newcommand{\M}{\mathcal{M}}
\newcommand{\F}{\mathcal{F}}
\newcommand{\K}{K_\varphi{}^\varphi}
\renewcommand{\S}{\mathcal{S}}
\newcommand{\Kf}{\mathcal{K}}
\newcommand{\V}{\mathcal{V}}
\newcommand{\A}{\mathcal{A}}
\newcommand{\C}{\mathcal{C}}

\newcommand{\two}[1]{{^{(2)}#1}}
\newcommand{\diff}{\textrm{d}}
\newcommand{\Lie}[1]{\mathcal{L}_{#1}}
\newcommand{\half}{\textstyle{\frac{1}{2}}}

\newcommand{\e}{\mathrm{e}}
\newcommand{\p}{\perp}
\newcommand{\n}{\parallel}
\renewcommand{\vec}{\mathbf}

\begin{document}

 \preprint{DAMTP-2005-17}

 \title{A strongly hyperbolic and regular reduction of\\ 
   Einstein's equations for axisymmetric spacetimes}

 \author{Oliver Rinne}
 \email{o.rinne@damtp.cam.ac.uk}
 \author{John M Stewart}
 \email{j.m.stewart@damtp.cam.ac.uk}
 \affiliation{Department of Applied Mathematics and Theoretical Physics,\\
 Centre for Mathematical Sciences, Wilberforce Road,\\
 Cambridge CB3 0WA, United Kingdom}
 \date{9 February 2005}

\begin{abstract}
This paper is concerned exclusively with axisymmetric spacetimes.
We want to develop reductions of Einstein's equations which are
suitable for numerical evolutions.
We first make a Kaluza-Klein type dimensional reduction followed by an
ADM reduction on the Lorentzian 3-space, the (2+1)+1 formalism.
We include also the Z4 extension of Einstein's equations adapted to
this formalism.
Our gauge choice is based on a generalized harmonic gauge condition.
We consider vacuum and perfect fluid sources.

We use these ingredients to construct a strongly hyperbolic 
first-order evolution system and exhibit its characteristic structure.
This enables us to construct constraint-preserving stable outer
boundary conditions.
We use cylindrical polar coordinates and so we provide a careful
discussion of the coordinate singularity  on axis.
By choosing our dependent variables appropriately we are able to
produce an evolution system in which each and every term is manifestly
regular on axis.
\end{abstract}



 \maketitle

\section{Introduction}
\label{sec:Intro}

In the early days of numerical relativity attention  focussed on
spherically symmetric spacetimes.
The next simplest subject would have been axisymmetric spacetimes, but very
little progress was  made in this area because, in polar
coordinates adapted to the symmetry, there is a coordinate
singularity on axis.  
Many attempts to deal with this proved unsuccessful, most evolutions
became unstable and attention quickly turned, with the rapidly
increasing capacity and speed of hardware, to spacetimes without symmetries. 
For a comprehensive review see e.g., \cite{Lehner}.
 
However axisymmetric situations arise frequently in astrophysics and
deserve further study.
The first question is, obviously, why were the evolutions unstable?
Suppose we assume that in a neighbourhood of the axis the spacetime is
regular so that we can introduce locally cartesian coordinates $(t, z,
x, y)$ with the axis given by $x=y=0$.
Then the Killing vector is $\xi = -y\partial/\partial x +
x\partial/\partial y$, and we say that $M$ is
\emph{axisymmetric} if ${\cal{L}}_\xi M = 0$.
We say that a tensor field $M$ is \emph{regular on axis} if its
components with respect to this chart  can be developed as a Taylor
series in $x$ and $y$ in a neighbourhood of the axis\footnote{For
  numerical purposes analytic and regular are synonymous.}.
We may introduce polar coordinates $r = \sqrt{x^2+y^2}$ and
$\varphi = \arctan(y/x)$, so that the Killing vector is 
$\xi = (\partial/\partial\varphi)$. 
Because the transformation between cartesian coordinates and polar
coordinates $(t, z, r, \varphi)$ is singular on axis\footnote{Note
  that the cartesian expression for the Killing vector $\xi$ vanishes
  on axis, whereas the polar form does not.}, axisymmetric
tensor fields which are regular on axis may take strange forms when
expressed in polar coordinates.
For example if $M_{\alpha\beta}$ is symmetric and has these properties
then
\begin{equation}
  \label{eq:introtensorreg}
  M_{\alpha\beta} = \left( \begin{array}{cccc} A & B & r D & r^2 F \\
      B & C & r E & r^2 G \\ r D & r E & H + r^2 J & r^3 K \\ 
      r^2 F & r^2 G & r^3 K & r^2 \left( H - r^2 J \right) \end{array} \right)
  \,,
\end{equation}
where $A,B,\ldots,K$ are functions of $t$, $z$ and $r^2$.
(See appendix \ref{sec:Conaxi} for this and related results.)
In particular both the metric and its first time-derivative must take
this form.  
If the evolution scheme fails to preserve precisely the indicated
$r$-dependencies then $M$ becomes irregular on axis and instability is
inevitable. 

Clearly the problems lie mainly in the last row and column of
\eqref{eq:introtensorreg}. 
In section \ref{sec:(2+1)+1} we outline a reduction of Einstein's
equations in spacetime ${\cal M}$ to a 3-dimensional Lorentzian
manifold ${\cal N}$ with chart $(t, z, r)$, effectively removing the
last row/column. 
(The relation between $M_{rr}$ and $M_{\varphi\varphi}$ still needs
careful treatment.)
Given a solution of the reduced equations in ${\cal N}$ one can
recover the solution in ${\cal M}$.
However there is no need to do this.
All axisymmetric quantities of physical interest in ${\cal M}$ have
their counterparts in ${\cal N}$, and so one might as well work in
${\cal N}$.
Because that manifold is Lorentzian we can perform an ADM decomposition.
The details of this $(2+1)+1$ scheme, with an arbitrary axisymmetric
matter source, are given in section \ref{sec:(2+1)+1}.

In the $(2+1)+1$ system we have a spatial foliation with lapse
function $\alpha$, shift vector $\eta^A$, and spatial 2-metric 
$H_{AB}$.  
(Here $A$, $B$, $\ldots$ range over $r$ and $z$.)
Since every 2-metric is conformally flat it is tempting to set 
$H_{rz}=0$, $H_{rr} = H_{zz} = \psi^4$ say.
The constraint equations imply elliptic equations for $\psi$ and the
shift vector components, and a quasi-maximal slicing condition
supplies an elliptic equation for $\alpha$.
Although elliptic equations can become computationally expensive,
multigrid techniques, if applicable,  can reduce significantly the cost.
The same choice was made by Choptuik et al \cite{CHLP}.
Like those authors we encountered several difficulties, the most
severe being that in near-critical collapse situations the discretized
version of the energy constraint equation for $\psi$ failed to remain
diagonally dominant so that the multigrid algorithm failed to
converge.
Garfinkle et al \cite{GD} used a conjugate gradient method which gets
round this particular problem.
If instead we used an evolution equation for $\psi$ there appeared to
be a serious violation of the energy constraint.

We decided therefore to look at the other extreme, to minimize the number
of elliptic equations solved.
There are many, apparently different, ways of doing this, but we chose
to implement the 
Z4 extension of Einstein's equation suggested  by Bona et al
\cite{Bona03}, \cite{Bona04}, but adapted to the (2+1)+1 scheme.
The Z(2+1)+1 scheme is outlined in section \ref{sec:Z(2+1)+1}.
There the energy and momentum constraints as well as the Geroch
constraint (which arises in the (2+1)+1 decomposition) are converted
into evolution equations for the new variables.

Pursuing this philosophy we would like to impose dynamical gauge
conditions, which are discussed in section \ref{sec:Gauge}.
These are generalizations of the harmonic gauge condition.
However cylindrical polar coordinates are not harmonic, and this means
that we can only do this for the lapse function.
For simplicity we assume that the shift vector vanishes.

We now have a system of pure evolution equations and it is highly
desirable to cast them into first order form.
This is carried out in section \ref{sec:For} where we express them in
conservation form with sources.
We show in section \ref{sec:Hype} that this system is \emph{strongly
  hyperbolic} and we write down the 
conversions between conserved and characteristic variables.
The characteristic speeds (divided by the lapse function $\alpha$) 
are $0$, $\pm 1$ and $\pm \sqrt{f}$ where $f$
is a constant parameter in the lapse evolution equation.
(The choice $f=1$ gives a harmonic time coordinate and a
\emph{symmetric hyperbolic} system.)

Knowing the characteristic structure enables us to formulate
constraint-preserving boundary conditions for the initial-boundary
value problem in section \ref{sec:Constr}.

In section \ref{sec:Reg} we return to our starting point, regularity
on axis. 
By a judicious choice of new dependent variables
(which does not affect the hyperbolicity) we
can write our first order strongly hyperbolic system in a form where
each and every term is manifestly regular on axis.

Finally we discuss some of the implementation possibilities and
applications of our scheme in section \ref{sec:Concl}.

%%% Local Variables: 
%%% mode: latex
%%% TeX-master: "master"
%%% End: 

\section{The (2+1)+1 formalism}
\label{sec:(2+1)+1}

Spacetime is assumed to be a four-dimensional manifold 
$(\M, g_{\alpha\beta})$ with signature $(-+++)$ and a preferred polar
coordinate chart $(t, r, z, \varphi)$. 
Axisymmetry means that there is an everywhere spacelike Killing
vector field $\xi = \partial / \partial \varphi$ with closed orbits.
The idea is to perform first a Kaluza-Klein-like reduction from $\M$ to
the Lorentzian three-dimensional manifold $\N$ formed by the trajectories 
of the Killing vector. 
(This was carried out for vacuum spacetimes by Geroch \cite{Geroch}.
Nakamura et al \cite{Maeda, Nakamura} extended the reduction to
include general matter sources and then considered the case of a
perfect fluid.
Choptuik et al \cite{CHLP} added a massless scalar field source,
albeit without rotation.
The current analysis includes arbitrary sources and rotation.)
After this an  ADM-like  reduction is  applied to $\N$.
We shall follow the notation of Nakamura et al. 
\cite{Maeda, Nakamura}, with some, clearly stated, changes.
In the following, Greek indices range over $t, r, z, \varphi$,
lower-case Latin indices over $t, r, z$, and upper-case Latin indices
over $r, z$.

Tensor fields $M^{\alpha...}{}_{\beta...}$ in $\M$ which are both
orthogonal to $\xi$ and are axisymmetric, i.e.,
\begin{equation*}
  M^{\alpha...}{}_{\beta...} \xi^{\beta} =  M^{\alpha...}{}_{\beta...}
  \xi_{\alpha} = \ldots = 0,\qquad
  \Lie{\xi} M^{\alpha...}{}_{\beta...} = 0
\end{equation*}
are said to be tensor fields \emph{in} $\N$.
Some basic tensor fields in $\N$ are
the norm of the Killing vector 
\begin{equation*}
  \lambda^2 = g_{\alpha\beta} \xi^\alpha \xi^\beta > 0 \, , 
\end{equation*}
the (Lorentzian) metric in $\N$,
\begin{equation*}
  h_{\alpha\beta} = g_{\alpha\beta} - \lambda^{-2}\xi_\alpha \xi_\beta \, , 
\end{equation*}
the Levi-Civita tensor
\begin{equation*}
  \epsilon_{\alpha\beta\gamma} =
  \lambda^{-1}\epsilon_{\alpha\beta\gamma\delta} \xi^\delta \, , 
\end{equation*}
and the ``twist vector''
\begin{equation}
  \label{eq:2.twist}
  \omega_\alpha = \epsilon_{\alpha\beta\gamma\delta} \xi^\beta \nabla^\gamma
  \xi^\delta \, , 
\end{equation}
where $\nabla$ is the covariant derivative of $g_{\alpha\beta}$.
%Note that neither $\xi^\alpha= \delta^\alpha{}_3$ nor 
%$\xi_\alpha = g_{\alpha 3}$ is in $\N$.
Note that neither $\xi^\alpha= \delta^\alpha{}_\varphi$ nor 
$\xi_\alpha = g_{\alpha \varphi}$ is in $\N$.
However if a solution of the equations in ${\cal N}$ is given then the
solution in ${\cal M}$ can be reconstructed \cite{Geroch}.
We shall argue that  a reconstruction is not necessary.
All physically relevant quantities in $\M$ have their counterparts in
$\N$.

We define the projections of the energy-momentum tensor as follows,
\begin{eqnarray}
  \label{eq:Tprojections}
  \tau &=& \lambda^{-2} \xi^\gamma \xi^\delta T_{\gamma\delta}\, , \nonumber\\
  \tau_\alpha &=& \lambda^{-2} h_\alpha{}^\gamma \xi^\delta 
  T_{\gamma\delta} \, ,  \\
  \tau_{\alpha\beta} &=& h_\alpha{}^\gamma h_\beta{}^\delta 
  T_{\gamma\delta} \, . \nonumber
\end{eqnarray}
Note that near the axis $\lambda = O(r)$, and the factors of $\lambda$
are included in \eqref{eq:Tprojections} to ensure that the projections
are either $O(1)$ or $O(r)$ near the axis, so that regularity
conditions (cf. section \ref{sec:Reg}) are easily imposed.

Following the standard ADM procedure, the three-dimensional manifold
$\N$ is foliated into two-dimensional  spacelike hypersurfaces
$\Sigma(t)$ with unit normal  $n_a = - \alpha \, \partial_a t$ and
intrinsic metric 
\begin{equation*}
  H_{ab} = h_{ab} + n_a n_b \, .
\end{equation*}
The line element in $\N$ takes the form
\begin{equation*}
  {\diff s}^2 = -\alpha^2 {\diff t}^2 + H_{AB} \left( \diff x^A +
  \eta^A \diff t \right) \left( \diff x^B + \eta^B \diff t \right) \, ,
\end{equation*}
where $\alpha$ is the lapse function and $\eta^A$ is the shift vector.
The extrinsic curvature is given by
\begin{equation*}
  \chi_{ab} = - H_a{}^c H_b{}^d n_{(c|d)} = -\half \Lie{n} H_{ab} \, ,
\end{equation*}
where $|$ denotes the covariant derivative compatible with $h_{ab}$.
The trace of the extrinsic curvature is abbreviated as $\chi = \chi_A{}^A$.
Further variables defined in each $\Sigma(t)$ are
the alternating symbol
 \begin{equation*}
  \epsilon_{AB} = n^c \epsilon_{cAB} \, ,
\end{equation*}
the $\varphi \varphi$-component of the extrinsic curvature,
\begin{equation*}
  \K = - \lambda^{-1} n^a \lambda_{,a} \, ,
\end{equation*}
and the rotational degrees of freedom,
\begin{eqnarray*}
  E^A &=& \lambda^{-3} \epsilon^{Ab} \omega_b \, ,  \nonumber\\
  B^\varphi &=& \lambda^{-3} n_a \omega^a \, . 
\end{eqnarray*}
Again, the last two definitions differ from those in \cite{Maeda,
  Nakamura} by factors of $\lambda$, and we write $B^\varphi$
  with an upstairs index. 
The various projections of the energy-momentum tensor are
\begin{eqnarray}
  \label{eq:tauprojections}
  J^\varphi &=& -n_a \tau^a \, , \nonumber\\
  S^A &=& H_a{}^A \tau^a \, ,  \nonumber\\
  \rho_H &=& n^a n^b \tau_{ab} \, , \\
  J_A &=& - H_A{}^a n^b \tau_{ab} \, ,  \nonumber\\
  S_{AB} &=& H_A{}^a H_B{}^b \tau_{ab} \, ,
\end{eqnarray}
and, of course $\tau$ defined in \eqref{eq:Tprojections}.

For convenience we also introduce
\begin{equation}
  \label{eq:AandL}
  A_A \equiv \alpha^{-1} \alpha_{,A} \, , \qquad
  L_A \equiv \lambda^{-1} \lambda_{,A} \, .
\end{equation}

We can now write down the projections of Einstein's equations
\begin{equation*}
  G_{\alpha\beta} = \kappa T_{\alpha\beta} \,,
\end{equation*}
which split into sets of elliptic constraint equations and hyperbolic
evolution equations.

The constraint equations are 
$\mathcal{C} = \mathcal{C}_A = \mathcal{C}_\varphi = 0$, where
the energy constraint is
\begin{equation}
  \label{eq:hamcons}
  \mathcal{C} \equiv \half (\chi^2 - \chi_{AB} \chi^{AB} + \two{R}) -
  L^A{}_{||A} - L_A L^A + \chi \K 
  - \frac{1}{4} \lambda^2 \left(E_A E^A + {B^\varphi}^2 \right) 
  - \kappa \rho_H  \, ,
\end{equation}
with $||$ denoting the covariant derivative of $H_{AB}$,
$\two{R_{AB}}$ the Ricci tensor of $H_{AB}$ and  $\two{R} =
\two{R_A{}^A}$ the Ricci scalar.
The linear momentum constraints are
\begin{equation}
  \label{eq:momcons}
  \mathcal{C}_A \equiv {\chi_A{}^B}_{|| B} - (\chi + \K)_{,A} 
  + L^B \chi_{AB} - L_A \K  
  - \half \lambda^2 B^\varphi \epsilon_{AB} E^B - \kappa J_A \, ,
\end{equation}
and the angular momentum or Geroch constraint is
\begin{equation}
  \label{eq:gercons}
  \mathcal{C}_\varphi \equiv \half E^A{}_{|| A} 
  + \frac{3}{2} L_A E^A - \kappa J^\varphi  \, .
\end{equation}

We shall express all the evolution equations in terms of the Lie
derivative along the normal direction, 
\begin{equation*}
  \Lie{n} = \alpha^{-1}( \partial_t - \Lie{\eta} ).
\end{equation*}
The time evolution of the metric variables is given by
\begin{eqnarray}
  \label{eq:d0H}
  \Lie{n} H_{AB} &=& - 2 \, \chi_{AB} \, ,\\
  \label{eq:d0lambda}
  \Lie{n} \lambda &=& - \lambda \K \, .
\end{eqnarray}
The evolution equations for the extrinsic curvature variables are
\begin{eqnarray}
  \label{eq:d0chi}
  \Lie{n} \chi_{AB} &=& \two{R}_{AB} + (\chi + \K) \chi_{AB} 
  - 2 \chi_A{}^C \chi_{CB} - A_{A||B} - A_A A_B
  - L_{A||B} - L_A L_B \nonumber\\ && 
  - \half \lambda^2 \left[ \epsilon_{AC} \epsilon_{BD} E^C E^D 
  - H_{AB} \left( E_C E^C - {B^\varphi}^2 \right) \right] \nonumber\\ && 
  - \kappa \left[ S_{AB} + \half
  H_{AB} \left( \rho_H - S_C{}^C - \tau \right) \right] \, , \\
  \label{eq:d0K} 
  \Lie{n} \K &=& 
  \K \left( \chi + \K \right)
  - L^A{}_{||A} - L_A (L^A + A^A) 
  - \half \lambda^2 \left(E_A E^A - {B^\varphi}^2 \right) \nonumber\\ && 
  - \half \kappa \left(\rho_H - S_A{}^A + \tau \right) \, .
\end{eqnarray}
The rotation variables evolve according to
\begin{eqnarray}
 \label{eq:d0E} 
  \Lie{n} E^A &=& (\chi + 3 \K) E^A + \epsilon^{AB} 
  B^\varphi{}_{,B}+ \epsilon^{AB} (A_B + 3 L_B) B^\varphi 
  - 2 \kappa S^A \, ,\\
  \label{eq:d0B}
  \Lie{n} B^\varphi &=& \chi B^\varphi + \epsilon^{AB} E_A A_B
  + \epsilon^{AB} E_{A || B} \, .
\end{eqnarray}
(The principal parts of these equations resemble the axisymmetric
Maxwell equations, which justifies the notation.  The Geroch
constraint \eqref{eq:gercons} is also a ``Maxwell equation''.)
Energy-momentum conservation $\nabla_\alpha T^{\alpha\beta} = 0$
implies the following evolution equations for the matter variables:
the energy conservation equation
\begin{equation}
  \label{eq:d0rhoH}
  \Lie{n} \rho_H = - J^A{}_{||A} - J^A ( 2 A_A + L_A ) + \chi \rho_H
  + \chi_{AB} S^{AB} + \K \tau + \lambda^2 E^A S_A \, ,
\end{equation}
the Euler equations
\begin{eqnarray}
  \label{eq:d0JA}
  \Lie{n} J_A &=& - S_{AB}{}^{||B} + J_A (\chi + \K) 
  - S_{AB} ( A^B + L^B) - A_A \rho_H + L_A \tau \nonumber\\&&
  + \lambda^2 (E_A J^\varphi + \epsilon_{AB} S^B B^\varphi) \, ,
\end{eqnarray}
and angular momentum conservation
\begin{equation}
  \label{eq:d0Jphi}
  \Lie{n} J^\varphi = - S^A{}_{||A} - S^A (A_A + 3 L_A) 
  + J^\varphi (\chi + 3 \K) \, .
\end{equation}

In many situations there may be a conserved particle number density
$N^\alpha$ satisfying $\Lie{\xi} N^\alpha = 0$. 
If we introduce the projections
\begin{eqnarray*}
  \nu^\alpha &=& h^\alpha{}_\beta N^\beta \,,\nonumber\\
  \Sigma^A &=& H^A{}_b \nu^b \,,\\
  \sigma &=& - \nu^a n_a \,,\nonumber
\end{eqnarray*}
then particle number conservation $\nabla_\alpha N^\alpha = 0$ implies
the following evolution equation for $\sigma$,
\begin{equation}
  \label{eq:d0sigma}
  \Lie{n} \sigma = - \Sigma^A{}_{||A} - \Sigma^A (A_A + L_A) 
    + \sigma (\chi + \K) \,.
\end{equation}

%%% Local Variables: 
%%% mode: latex
%%% TeX-master: "master"
%%% End: 

\section{The Z4 extension}
\label{sec:Z(2+1)+1}

Bona et al. \cite{Bona03} suggested the following covariant extension of
the Einstein field equations:
\begin{equation}
  \label{eq:Z4}
  R_{\alpha\beta} + 2 \nabla_{(\alpha} Z_{\beta)} = 
  \kappa \left( T_{\alpha\beta} - \half T g_{\alpha\beta} \right) \,,
\end{equation}
which reduces to the Einstein equations if and only if $Z_\alpha = 0$.

We shall assume that $Z_\alpha$ shares the  axisymmetry, i.e., 
$\Lie{\xi} Z_\alpha = 0$.
To apply the (2+1)+1 formalism to equations \eqref{eq:Z4}, it is
convenient to rewrite them as Einstein's equations $G_{\alpha\beta} =
\kappa \widetilde T_{\alpha\beta}$ with a modified energy-momentum tensor
\begin{equation}
  \label{eq:Ttilde}
  \widetilde T_{\alpha\beta} = T_{\alpha\beta} - \frac{2}{\kappa} \left( 
    \nabla_{(\alpha} Z_{\beta)} - \half g_{\alpha\beta}\nabla_\gamma Z^\gamma 
    \right) \, .
\end{equation}

The $Z$ covector is decomposed into parts parallel and orthogonal
to the Killing vector $\xi^\alpha$,
\begin{equation*}
  Z^\varphi = \lambda^{-2} \xi^\alpha Z_\alpha \,, \qquad 
  Z_a = h_a{}^\alpha Z_\alpha \,,
\end{equation*}
and further parallel and orthogonal to the timelike normal $n_a$,
\begin{equation*}
  \theta = - n^a Z_a \,, \qquad
  Z_A = H_A{}^a Z_a \,.
\end{equation*}

We now compute the modified matter variables corresponding to 
$\widetilde T_{\alpha\beta}$, finding

\begin{eqnarray}
  \label{eq:modmat}
  \widetilde \tau &=& \tau + \kappa^{-1} \left[ \Lie{n} \theta 
  + Z^A{}_{||A} + (A_A - L_A ) Z^A + (\K - \chi) \theta \right] \,, \nonumber\\
  \widetilde S_A &=& S_A + \kappa^{-1} \left[ - Z^\varphi{}_{,A} 
    + B^\varphi \epsilon_{AB} Z^B + E_A \theta  \right]
  \,, \nonumber\\
  \widetilde J^\varphi &=& J^\varphi + \kappa^{-1} \left[ \Lie{n} Z^\varphi 
    + E^A Z_A \right] \,, \nonumber\\
  \widetilde S_{AB} &=& S_{AB} + \kappa^{-1} \big[ -2 Z_{(A||B)} 
    + 2 \chi_{AB} \theta  \\ &&  \qquad \qquad \qquad
    + H_{AB} \left\{ \Lie{n} \theta  + Z^C{}_{||C} + (A_C + L_C) Z^C
      - (\chi + \K) \theta \right\} \big] \,, \nonumber\\
  \widetilde J_A &=& J_A + \kappa^{-1} \left[ \Lie{n} Z_A - \theta_{,A} 
    + 2 \chi_{AB} Z^B + A_A \theta  \right] \, , \nonumber\\
  \widetilde \rho_H &=& \rho_H + \kappa^{-1} \left[ \Lie{n} \theta - Z^A{}_{||A}
    + (A_A - L_A) Z^A + (\chi + \K) \theta \right] \, .\nonumber
\end{eqnarray}

Inserting the above into the standard (2+1)+1 equations, we obtain
the Z(2+1)+1 equations. 
Because of the Lie derivatives in \eqref{eq:modmat}, the constraints
(\ref{eq:hamcons}--\ref{eq:gercons}) become evolution equations 
for the Z covector,
\begin{eqnarray}
  \label{eq:d0theta}
  \Lie{n} \theta &=& \mathcal{C} + (L_A - A_A) Z^A 
  + Z^A{}_{||A} - (\chi + \K) \theta \,,\\
  \label{eq:d0ZA}
  \Lie{n} Z_A &=& \mathcal{C}_A
  - 2 \chi_{AB} Z^B - A_A \theta  + \theta_{,A} \,,\\
  \label{eq:d0Zphi}
  \Lie{n} Z^\varphi &=& \mathcal{C}_\varphi - E^A Z_A \, . 
\end{eqnarray}

The main evolution equations are modified as follows,
\begin{eqnarray*}
  \Lie{n} \chi_{AB} &=& \ldots + 2 Z_{(A||B)} - 2 \chi_{AB} \theta \,,\\
  \Lie{n} \K &=& \ldots + 2 L_A Z^A - 2 \K
    \theta \,,\\
  \Lie{n} E^A &=& \ldots  + 2 Z^\varphi{}^{,A} - 2 E^A \theta 
    - 2 B^\varphi \epsilon^{AB} Z_B \,,
\end{eqnarray*}
where $\ldots$ denote the right-hand-sides of \eqref{eq:d0chi},
\eqref{eq:d0K} and \eqref{eq:d0E}, respectively.
The remaining evolution equations are unchanged.

%%% Local Variables: 
%%% mode: latex
%%% TeX-master: "master"
%%% End: 

\newpage
\section{Dynamical gauge conditions}
\label{sec:Gauge}

To complete our evolution formalism, we need to prescribe the gauge
variables $\alpha$ and $\eta^A$. 
It is well-known \cite{DeDonder, Fock}
that the Einstein equations can be written in symmetric hyperbolic form
\begin{equation*}
   g^{\gamma\delta}g_{\alpha\beta,\gamma\delta} \simeq 0 
\end{equation*}
($\simeq$ denoting equality to principal parts)
by adopting harmonic gauge, which for the Z4 extension of the
field equations becomes
\begin{equation*}
  g^{\gamma\delta}x^\alpha{}_{;\gamma\delta} = 2 Z^\alpha \, ,
\end{equation*}
where we are treating $x^\alpha$ as a scalar field.
It would therefore be desirable to choose this or a related gauge for
the system at hand. 
However, these gauge conditions are incompatible with cylindrical
polar coordinates; the wave operator contains a term
$r^{-1} \partial_r$, and the
right-hand-side of the evolution equation for the shift vector
becomes singular on the axis. 
However, we can keep the time component of the harmonic gauge
equation, which in (2+1)+1 language reads
\begin{equation*}
  \partial_t \alpha = - \alpha^2 ( \chi + \K - 2 \theta) \, .
\end{equation*}
This condition has been generalized to \cite{Bona04}
\begin{equation}
  \label{eq:d0alpha}
  \partial_t \alpha = - f \alpha^2 ( \chi + \K - m \theta) \, ,
\end{equation}
where $f$ and $m$ are free constant parameters.

We choose the shift vector to vanish: $\eta^A = 0$. Apart from
simplifying the evolution equations, this choice also makes it
much easier to set up outer boundary conditions because a variable shift
would mean that the characteristic speeds could change sign at the
boundary (see section \ref{sec:Hype}).
Our choice for the shift vector means that henceforth
\begin{equation*}
  \Lie{n} = \alpha^{-1} \partial_t \,.
\end{equation*}

%%% Local Variables: 
%%% mode: latex
%%% TeX-master: "master"
%%% End: 

\section{First-order reduction}
\label{sec:For}

The Z(2+1)+1 equations are first-order in time but second-order in
space. In order to analyze their hyperbolicity and to be able to treat them
with standard numerical methods for hyperbolic conservation laws, we perform a
first-order reduction. We focus on the geometry evolution equations here; 
for given geometry, the matter evolution equations form a separate 
strongly hyperbolic system (see appendix \ref{sec:matter}).

To eliminate the second spatial derivatives, we introduce new
variables for the first spatial derivatives of the two-metric,
\begin{equation*}
  D_{ABC} \equiv \half \partial_A H_{BC} \,,
\end{equation*}
and we regard $A_A$ and $L_A$ defined by (\ref{eq:AandL}) 
as independent variables. 
Evolution equations for these new first-order variables can
be obtained from (\ref{eq:d0H}), (\ref{eq:d0lambda}) and (\ref{eq:d0alpha})
by commuting space and time derivatives:
\begin{eqnarray}
  \label{eq:d0D}
  \Lie{n} D_{ABC} &=& - \chi_{BC ,A} \,,\\ 
  \label{eq:d0L}
  \Lie{n} L_A &=& - {\K}_{,A} \,,\\
  \label{eq:d0A}
  \Lie{n} A_A &=& - f (\chi_{,A} + {\K}_{,A} - m \theta_{,A}) \,.
\end{eqnarray}

The two independent traces of $D_{ABC}$ are denoted by
\begin{equation*}
  D^I{}_A \equiv D_{AB}{}^B \,, \qquad 
  D^{II}{}_A \equiv D^B{}_{BA} \, ,
\end{equation*}
where indices are (formally, for $D$ is not a tensor) raised and lowered 
with the 2-metric $H_{AB}$.
The De Donder-Fock decomposition \cite{DeDonder, Fock} 
is used for the Ricci tensor,
\begin{eqnarray*}
  \two{R}_{AB} &=& - D^C{}_{AB,C} + 2 D^{II}{}_{(A,B)} - D^I{}_{(A,B)} 
  - 2 D_{CAB} D^{II\,C}  \nonumber \\ &&
  - \Gamma_{CAB} ( 2 D^{II\,C} - D^{I\,C} ) 
  + 4 D_{CDA} D^{CD}{}_B - \Gamma_{ACD} \Gamma_B{}^{CD} \,,
\end{eqnarray*}
where of course the Christoffel symbols are given by
\begin{equation}
  \label{eq:Chris}
  \Gamma_{ABC} = D_{CAB} + D_{BCA} - D_{ABC}.
\end{equation}

It is then straightforward to write the Z(2+1)+1 equations in
\emph{conservation form}
\begin{equation}
  \label{eq:consform}
  \partial_t \vec u + \left[ \alpha \vec \F^D (\vec u) \right]_{,D} 
  = \alpha \vec \S (\vec u) \,.
\end{equation}
Here, the set of conserved variables is
\begin{equation*}
  \vec u = (H_{AB}, \lambda, \alpha, D_{ABC}, L_A, A_A, \chi_{AB},
  \K, E^A, B^\varphi, \theta, Z_A, Z^\varphi)^T \,,
\end{equation*}
and the components of the flux vector are given by
\begin{eqnarray*}
  \F^D{}_{H_{AB}} &=& 0 \, ,\\
  \F^D{}_{\lambda} &=& 0 \, ,\\
  \F^D{}_{\alpha} &=& 0 \, ,\\
  \F^D{}_{D_{ABC}} &=& \delta_A^D \chi_{BC} \, ,\\
  \F^D{}_{L_A} &=& \delta_A^D \K \, ,\\
  \F^D{}_{A_A} &=& \delta_A^D f (\chi + \K - m \theta) \,,\\
  \F^D{}_{\chi_{AB}} &=& D^D{}_{AB} - \delta^D_{(A} \left( 
    2 D^{II}{}_{B)} + 2 Z_{B)} -  D^{I}{}_{B)} - L_{B)} - A_{B)} \right) \,,\\
  \F^D{}_{\K} &=& L^D \,,\\
  \F^D{}_{E^A} &=& - 2 H^{AD} Z^\varphi - \epsilon^{AD} B^\varphi \,,\\
  \F^D{}_{B^\varphi} &=& -\epsilon^{AD} E_A \,,\\
  \F^D{}_{\theta} &=& D^{I\,D} - D^{II\,D} + L^D - Z^D \,,\\
  \F^D{}_{Z_A} &=& -\chi_A{}^D + \delta_A^D ( \chi + \K 
    - \theta ) \,,\\
  \F^D{}_{Z^\varphi} &=& - \half E^D \,.
\end{eqnarray*}
$\S(\vec u)$ is a source term containing no derivatives,
\begin{eqnarray*}
  \S_{H_{AB}} &=& -2 \chi_{AB} \,,\\
  \S_\lambda &=& -\lambda \K \,,\\
  \S_\alpha &=& -f \alpha (\chi + \K - m \theta) \,,\\
  \S_{D_{ABC}} &=& 0 \,,\\
  \S_{L_A} &=& 0 \,,\\
  \S_{A_A} &=& 0 \,,\\
  \S_{\chi_{AB}} &=& A_C D^C{}_{AB} + A_{(A} \left( -2 D^{II}{}_{B)} +
     D^I{}_{B)} + L_{B)} + A_{B)} - 2 Z_{B)} \right) \\&& 
     - L_A L_B - A_A A_B 
     + \Gamma^C{}_{AB} (L_C + A_C) + \chi_{AB} (\chi + \K) 
     - 2 \chi_A{}^C \chi_{CB} \\&&  
     - \half \lambda^2 \left[ \epsilon_{AC} \epsilon_{BD} E^C E^D 
       - H_{AB} (E_C E^C - B^\varphi{}^2 ) \right] \\&&
     - \kappa \left[ S_{AB} + \half H_{AB} (\rho_H - S_C{}^C - \tau)
     \right] - \Gamma_{CAB} (2 Z^C + 2 D^{II\,C} - D^{I\,C}) \\&& 
     - 2 \theta \chi_{AB} - 2 D_{CAB} D^{II\,C} + 4 D_{CDA} D^{CD}{}_B 
     - \Gamma_{ACD} \Gamma_B{}^{CD} \,,\\
  \S_{\K} &=& \K (\chi + \K)
     + L_A (2 Z^A - L^A - D^{I\,A}) - \half \lambda^2 (E_A E^A + B^\varphi{}^2)
      - 2 \K \theta \\&& 
      - \half \kappa (\rho_H - S_C{}^C + \tau) \,,\\
  \S_{E^A} &=& (\chi + 3 \K - 2 \theta) E^A +
     \epsilon^{AB} B^\varphi (3 L_B - 2 Z_B + D^I{}_B) - 2 \kappa S^A 
      + (4 D^{II\,A} - 2 A^A) Z^\varphi \,,\\
  \S_{B^\varphi} &=& \chi B^\varphi + \epsilon^{AB} E_A D^I{}_B \,,\\
  \S_\theta &=& A_A (D^{I\,A} - D^{II\,A} + L^A - Z^A) +
     D_{ABC} D^{ABC} - \half \Gamma_{ABC} \Gamma^{ABC} 
     - \half D^I{}_A D^{I\,A} \\&& - L_A (L^A + D^{I\,A}) 
     + \half (\chi^2 - \chi_{AB} \chi^{AB}) + \chi \K
     -\frac{1}{4} \lambda^2 (E_A E^A + B^\varphi{}^2) 
     - \kappa \rho_H \\&&
     + (L_A - A_A + D^I{}_A) Z^A - (\chi + \K) \theta \,,\\
  \S_{Z_A} &=& - A_B \chi_A{}^B + A_A (\chi + \K - \theta) 
     + \chi_{AB} ( D^{I\,B} + L^B - 2 Z^B) 
     - \Gamma^C{}_{AB} \chi_C{}^B  \\&& - L_A \K - \half \lambda^2 B^\varphi
     \epsilon_{AB} E^B - \kappa J_A - A_A \theta \,,\\
  \S_{Z^\varphi} &=& \half E^A (D^I{}_A + 3 L_A - 2 Z_A - A_A) -
    \kappa J^\varphi \,.
\end{eqnarray*}

Note that $H_{AB}$, $\lambda$ and $\alpha$ have vanishing fluxes and
thus trivially propagate along the normal direction. The rotation
variables $(E^A, B^\varphi, Z^\varphi)$ form a decoupled subsystem
analogous to Maxwell's equations.

%%% Local Variables: 
%%% mode: latex
%%% TeX-master: "master"
%%% End: 

\section{Hyperbolicity}
\label{sec:Hype}

To investigate the hyperbolicity of the Z(2+1)+1 system, we pick a unit
covector $\mu_A$ and define an orthogonal covector 
\begin{equation}
  \label{eq:pdef}
  \pi_A = \epsilon_{AB} \mu^B \,,
\end{equation}
so that
\begin{equation*}
  \mu_A \mu^A = \pi_A \pi^A = 1 \,, \qquad 
  \mu_A \pi^A = 0 \,.
\end{equation*}
Thus $(\mu^A, \pi^A)$ form an orthonormal basis for $T(\Sigma)$.
Projection along $\mu$ and $\pi$ is denoted as
\begin{equation}
  \label{eq:projnotation}
  V^\parallel \equiv V^A \mu_A \,, \qquad V^\perp \equiv V^A \pi_A \,.
\end{equation}
It can be verified that the Jacobian matrix
\begin{equation*}
  B^\parallel \equiv \frac{\partial \vec \F^\parallel}{\partial \vec u}
\end{equation*}
corresponding to the flux in the $\mu$-direction is real,
diagonalizable and has  complete sets of left and right eigenvectors
for arbitrary $\mu_A$, i.e., the system is \emph{strongly hyperbolic}.
The characteristic speeds are\footnote{Note that a factor of $\alpha$
  has been taken out of the fluxes (\ref{eq:consform}). 
  Note also that for a nonzero shift vector 
  $\eta^A$, a term $- \alpha^{-1} \eta^\parallel$ would have to be added to
  the above characteristic speeds.}
\begin{eqnarray*}
  \lambda_0 &=& 0 \,,\\
  \lambda_1^\pm &=& \pm 1 \quad \textrm{(speed of light)} \,,\\
  \lambda_f^\pm &=& \pm \sqrt{f} \quad \textrm{(gauge speed)}\,.
\end{eqnarray*}
For $f \leqslant 1$, all characteristic speeds are causal, for $f = 1$
(harmonic slicing) they are all physical.

The characteristic variables (left eigenvectors dotted into the
conserved variables) are given by
\begin{itemize}
  \item Normal modes (eigenvalue $0$)
  \begin{eqnarray*}
    l_{0,1} &=& f m (D_{\n\p\p} + L_\n - D_{\p\p\n} - Z_\n) 
              - f (D_{\n\n\n} + D_{\n\p\p} + L_\n) + A_\n \,,\\
    l_{0,2} &=& f m (D_{\p\n\n} + L_\p - D_{\n\n\p} - Z_\p)
              - f (D_{\p\n\n} + D_{\p\p\p} + L_\p) + A_\p \,,\\
    l_{0,3} &=& D_{\p\n\n} \,,\\
    l_{0,4} &=& D_{\p\p\n} \,,\\
    l_{0,5} &=& D_{\p\p\p} \,,\\
    l_{0,6} &=& L_\p \,,\\
    l_{0,7} &=& A_\p \,,
  \end{eqnarray*}
  along with the zeroth-order variables $H_{AB}$, $\lambda$ and $\alpha$.
  \item Light cone modes (eigenvalue $\pm 1$)
  \begin{eqnarray*}
    l_{1,1}^\pm &=& \theta \pm (D_{\n\p\p} + L_\n - D_{\p\p\n} - Z_\n) \,,\\
    l_{1,2}^\pm &=& \chi_{\n\p} \pm \half ( A_\p + D_{\p\n\n} -
                    D_{\p\p\p} + L_\p - 2 Z_\p) \,,\\
    l_{1,3}^\pm &=& \chi_{\p\p} \pm D_{\n\p\p} \,,\\
    l_{1,4}^\pm &=& \K \pm L_\n \,,\\
    l_{1,5}^\pm &=& E^\n \mp 2 Z^\varphi \,,\\
    l_{1,6}^\pm &=& E^\p \mp B^\varphi \,,
  \end{eqnarray*}
  \item Lapse cone modes (eigenvalue $\pm \sqrt{f}$)
  \begin{eqnarray}
    \label{eq:lapsecone}
    l_f^\pm &=& A_\n - \frac{f(m-2)}{f-1} (D_{\n\p\p} + L_\n -
    D_{\p\p\n} - Z_\n) \nonumber\\ &&
    \pm \sqrt{f} \left[ \chi_{\n\n} + \chi_{\p\p} + \K 
    - \left( \frac{f(m-2)}{f-1} + 2 \right) \theta \right] \, .
  \end{eqnarray}  
\end{itemize}
The inverse transformation from characteristic to conserved variables
is given in Appendix \ref{sec:InvTrafo}.

In the special case $f=1$, we must set $m=2$ to enforce strong
hyperbolicity, and the singular term $(m-2)/(f-1)$ in 
(\ref{eq:lapsecone}) is to be replaced with an arbitrary fixed
constant (e.g. $0$ for simplicity).
In this case, the system is even \emph{symmetric hyperbolic},
i.e. $B^\parallel$ is symmetrizable with a symmetrizer that is independent of
the direction $\mu_A$. This is not surprising because the choice of
parameters $f=1, \, m=2$ in (\ref{eq:d0alpha}) corresponds to 
harmonic slicing. An energy for the system is given by
\begin{eqnarray*}
  \mathcal{E} &=& \chi_{AB} \chi^{AB} + \lambda_{ABC} \lambda^{ABC}
  + (\K + \chi - 2 \theta)^2 + A_A A^A \nonumber\\
  && + V_A V^A + {\K}^2 + L_A L^A + E_A E^A + {B^\varphi}^2
  + 4 {Z^\varphi}^2 \,,
\end{eqnarray*}
where
\begin{eqnarray*}
  V_A &\equiv& A_A + D^I{}_A + L_A - 2 D^{II}{}_A - 2 Z_A \,,\nonumber\\
  \lambda^A{}_{BC} &\equiv& D^A{}_{BC} + \delta^A_{(B} V_{C)} \,.
\end{eqnarray*}
$\mathcal{E}$ is positive definite and its time derivative is
(considering principal parts only) a total divergence.

%%% Local Variables: 
%%% mode: latex
%%% TeX-master: "master"
%%% End: 

\section{Constraint-preserving boundary conditions}
\label{sec:Constr}

In the Z(2+1)+1 formalism, the standard (2+1)+1 constraints
(\ref{eq:hamcons}--\ref{eq:gercons}) are replaced with the
\emph{algebraic} constraints 
\begin{equation} 
   \label{eq:Zconstr}
   \theta = Z_A = Z^\varphi = 0 \, .
\end{equation}
If those hold at all times, the (2+1)+1 constraints are
automatically satisfied by virtue of the evolution equations for the
$Z$ vector (\ref{eq:d0theta}--\ref{eq:d0Zphi}). 
On the initial time level, one imposes \eqref{eq:Zconstr} and solves
the (2+1)+1 constraints (\ref{eq:hamcons}--\ref{eq:gercons})
so that both the $Z$ vector and its time
derivative vanish. In addition, the outer boundary conditions have to be
chosen so that no constraint-violating modes propagate into the
computational domain at any time.

A propagation system for (\ref{eq:Zconstr}) follows from
the contracted Bianchi identities, which imply 
$\nabla_\alpha \tilde T^{\alpha \beta} = 0$.
If we impose the standard matter evolution equations 
$\nabla_\alpha T^{\alpha \beta} = 0$ in addition, (\ref{eq:Ttilde})
yields the following homogeneous wave equation for the $Z$ covector:
\begin{equation}
  \label{eq:BoxZ}
  \nabla_\beta\nabla^\beta Z_\alpha + R_{\alpha \beta} Z^\beta = 0 \, .
\end{equation}

To obtain the (2+1)+1 reduction of \eqref{eq:BoxZ}, 
we take an additional time derivative 
of the evolution equations for the $Z$ vector 
(\ref{eq:d0theta}--\ref{eq:d0Zphi}). To principal parts, we have
\begin{eqnarray}
  \label{eq:subsidiary}
  \Lie{n}^2 \theta - \partial_B \partial^B \theta &\simeq& 0 \,,\nonumber\\
  \Lie{n}^2 Z_A - \partial_B \partial^B Z_A &\simeq&
    \half \partial_B \left[ \partial_A (A^B + D^{I\, B} + L^B)
      - \partial^B ( A_A + D_{I\, A} + L_A ) \right. \nonumber\\ &&
      \left. \qquad -2 ( \partial_C D^{BC}{}_A - \partial^B D^{II}{}_A)
      \right] \,,\\
  \Lie{n}^2 Z^\varphi - \partial_B \partial^B Z^\varphi &\simeq& 0
    \,.\nonumber
\end{eqnarray} 
Note that the right-hand-sides are zero if the \emph{ordering constraints}
\begin{eqnarray}
  \label{eq:orderingconstr}
  \partial_D D_{ABC} &=& \partial_A D_{DBC} \,,\nonumber\\
  \partial_B L_A &=& \partial_A L_B \,,\\
  \partial_B A_A &=& \partial_A A_B \nonumber
\end{eqnarray}
are satisfied, which corresponds to commuting of partial derivatives 
in the definition of the first-order variables.  
Although this holds analytically, it may not obtain numerically, and
so the terms have to be included in the subsidiary system
\eqref{eq:subsidiary}.

We choose an orthonormal basis $\mu^A, \pi^A$ such that $\mu^A$ is
normal to the outer boundary under consideration. Indices contracted
with $\mu^A$ ($\pi^A$) are denoted by $\n$ ($\p$).
Equations \eqref{eq:subsidiary} decompose into
\begin{eqnarray}
  \Lie{n}^2 \theta - \partial_\n^2 \theta - \partial_\p^2 \theta
    &\simeq& 0 \,,\nonumber\\
  \Lie{n}^2 Z_\n - \partial_\n^2 Z_\n - \partial_\p^2 Z_\n &\simeq&
    \partial_\n^2 D_{\p\p\n} \nonumber\\ &&
    + \half \partial_\n \partial_\p (A_\p - D_{\p\n\n }+ D_{\p\p\p} 
      - 2 D_{\n\n\p} + L_\p) \nonumber\\ &&
    + \half \partial_\p^2 (-A_\n + D_{\n\n\n} - D_{\n\p\p} - L_\n) \,,
    \nonumber\\
  \Lie{n}^2 Z_\p - \partial_\n^2 Z_\p - \partial_\p^2 Z_\p &\simeq&
    \half \partial_\n^2 (-A_\p - D_{\p\n\n} + D_{\p\p\p} - L_\p) \nonumber\\
    && + \half \partial_\n \partial_\p (A_\n + D_{\n\n\n} - D_{\n\p\p} 
    - 2 D_{\p\p\n} + L_\n ) \nonumber\\
    && + \partial_\p^2 D_{\n\n\p}  \,,\nonumber\\
  \Lie{n}^2 Z^\varphi - \partial_\n^2 Z^\varphi - \partial_\p^2 Z_\varphi
    &\simeq& 0 \,.\nonumber
\end{eqnarray} 
This second-order system can be written in first-order form 
by introducing new variables for the temporal and spatial
derivatives of the fundamental variables. Absorbing boundary
conditions can then be imposed in the $\n$ direction, which read 
(after replacing the new variables with their definitions as first
derivatives of the fundamental variables)
\begin{eqnarray}
  \label{eq:cpbctheta}
  \Lie{n} \theta &\doteq& - \partial_\n \theta  \,,\\
  \label{eq:cpbczpar}
  \Lie{n} Z_\n &\doteq& - \partial_\n (Z_\n + D_{\p\p\n})
    - \half \partial_\p ( A_\p - D_{\p\n\n} + D_{\p\p\p} 
    - 2 D_{\n\n\p} + L_\p ) \,,\\
  \label{eq:cpbczperp}  
  \Lie{n} Z_\p &\doteq& -\partial_\n \left[ Z_\p - \half (A_\p + D_{\p\n\n} -
    D_{\p\p\p} + L_\p) \right] \nonumber \\ 
    &&- \half \partial_\p (A_\n + D_{\n\n\n} - D_{\n\p\p} - 2
    D_{\p\p\n} + L_\n) \,,\\
  \label{eq:cpbczphi}  
  \Lie{n} Z^\varphi &\doteq& -\partial_\n Z^\varphi \,, 
\end{eqnarray}
where $\doteq$ denotes equality at the boundary.

The straightforward way of setting up stable outer boundary conditions 
for the \emph{main} evolution system is to set all ingoing modes to 
zero at the outer 
boundaries of the computational domain (so-called 
\emph{absorbing} boundary conditions).
For a symmetric hyperbolic system, this leads to a
well-posed initial boundary value problem (IBVP) \cite{Rauch, Secchi}.
Our goal is to investigate under which circumstances such boundary
conditions \emph{imply} the constraint-preserving first-order
conditions (\ref{eq:cpbctheta}--\ref{eq:cpbczphi}).
Suppose $l^-$ is an incoming mode with characteristic speed $-\lambda$
($\lambda > 0$).
Setting $l^-$ to zero at the entire boundary at all times implies that
\begin{equation}
  \label{eq:cond1}
  0 \doteq \partial_\p l^- \equiv L_1(l^-)
\end{equation}
and 
\begin{equation}
  \label{eq:cond2}
  0 \doteq \Lie{n} \, l^- \simeq
  \lambda \, \partial_\n l^-{} - \partial_\p \F^\p{}_{l^-} \equiv L_2 (l^-)\,,
\end{equation}
where $\doteq$ denotes equality at the boundary and we have used the
general general evolution equation for $l^-$.
The relations \eqref{eq:cond1} and \eqref{eq:cond2} will now be used to
manipulate the general evolution equations for the $Z$ vector at the
boundary. As mentioned in section \ref{sec:Hype}, the evolution system
decouples into a non-rotational and a rotational part at the level of
principal parts, and we consider the two subsystems separately.

Let us start with the general evolution equations for $\theta$ and
$Z_A$, decomposed with respect to our boundary-adapted basis:
\begin{eqnarray}
  \Lie{n} \theta &\simeq& - \partial_\n (D_{\n\p\p} - D_{\p\p\n} + L_\n - Z_\n)
        - \partial_\p (D_{\p\n\n} - D_{\n\n\p} + L_\p - Z_\p) \,,\nonumber\\
  \Lie{n} Z_\n &\simeq& - \partial_\n (\chi_{\p\p} + \K - \theta) +
        \partial_\p \chi_{\n\p} \,,\nonumber\\
  \Lie{n} Z_\p &\simeq& \partial_\n \chi_{\n\p} - \partial_\p 
       (\chi_{\n\n} + \K -  \theta) \,.\nonumber
\end{eqnarray}
If we add to these suitable combinations of equations \eqref{eq:cond1} and
\eqref{eq:cond2},
\begin{eqnarray}
  \Lie{n} \theta &+=& L_2(-l^-_{1,1}) \,,\nonumber\\
  \Lie{n} Z_\n &+=& L_2(-l^-_{1,1} + l^-_{1,3} + l^-_{1,4}) 
                  + L_1(-l^-_{1,2}) \,,\nonumber\\
  \Lie{n} Z_\p &+=& L_2(-l^-_{1,2}) 
                  + L_1(l^-_{1,1} - l^-_{1,3} - l^-_f) \,,\nonumber
\end{eqnarray}
we obtain precisely the constraint-preserving boundary conditions
(\ref{eq:cpbctheta}--\ref{eq:cpbczperp}), \emph{provided that we choose
harmonic gauge} $f = 1,\, m = 2, \frac{m-2}{f-1} = 0$. 
(For different values of $f$ and $m$,
the derivatives tangential to the boundary in the evolution equation
for $Z_\p$ differ from the constraint-preserving boundary condition 
\eqref{eq:cpbczperp}.)
Hence absorbing boundary conditions for the non-rotational part of the
evolution system preserve the constraints.

However, this is not true for the rotational subsystem. A little
calculation shows that the only dissipative boundary condition
consistent with \eqref{eq:cpbczphi} is 
\begin{equation*}
 l^-_{1,5} \doteq 0, \quad 
 l^-_{1,6} \doteq l^+_{1,6} \Leftrightarrow B^\varphi \doteq 0 \,.
\end{equation*}
Fortunately, there is a much simpler solution to this problem: 
we delete $Z^\varphi$ entirely from the evolution system. 
Note that setting $Z^\varphi \equiv 0$ does not break
general covariance because our spacetime has a Killing vector and we
simply choose $Z_\alpha \xi^\alpha = 0$.
The remaining rotational subsystem is still symmetric hyperbolic, 
and the constraint
\begin{equation*}
  \C_\varphi \simeq \half E^A{}_{,A}
\end{equation*}
has zero speed with respect to the boundary:
\begin{equation*}
  \Lie{n} \C_\varphi \simeq \half \epsilon^{AB}
  B^\varphi{}_{,AB} = 0 \,.
\end{equation*}
The absorbing boundary condition
\begin{equation}
  \label{eq:rotabsbc}
  l^-_{1,6} = E^\p + B^\varphi \doteq 0
\end{equation}
leads to a well-posed IBVP for this system.

In addition to the $Z$ constraints, there are differential constraints 
related to the definition of the first-order variables,
\begin{eqnarray*}
  \label{eq:diffconstr}
  D_{ABC} &=& \half \partial_A H_{BC} \,,\nonumber\\
  L_A &=& \lambda^{-1} \partial_A \lambda \,,\\
  A_A &=& \alpha^{-1} \partial_A \alpha  \,,\nonumber
\end{eqnarray*}
and the ordering constraints \eqref{eq:orderingconstr}.
By virtue of the evolution equations 
(\ref{eq:d0H}, \ref{eq:d0lambda}, \ref{eq:d0alpha})
and (\ref{eq:d0D}, \ref{eq:d0L}, \ref{eq:d0A}), 
it is easy to see that the differential and ordering constraints have
zero speed with respect to the boundary.

In summary, we have shown that for harmonic gauge, absorbing boundary
conditions for the non-rotational part of the main evolution system
are automatically constraint-preserving, and that there are two ways
of setting up constraint-preserving stable outer boundary
conditions for the rotational subsystem. We thus have a well-posed IBVP with
constraint-preserving boundary conditions. 
The result concerning the non-rotational part carries over immediately 
to the general Z4 system without spacetime symmetries.
Constraint-preserving boundary conditions for the Z4 system have
recently also been analysed by Bona et al. \cite{Bona04a}: in essence, 
those authors propose to implement the constraint-preserving boundary 
conditions (\ref{eq:cpbctheta}--\ref{eq:cpbczperp}) directly. 
While it is not clear analytically that such boundary conditions lead
to a well-posed IBVP, Bona et al.~are able to prove a necessary condition 
and to demonstrate numerical stability.

%%% Local Variables: 
%%% mode: latex
%%% TeX-master: "master"
%%% End: 

\section{Regularity on axis}
\label{sec:Reg}

Extra care is needed in axisymmetric situations because the coordinate
system that is adapted to the symmetry, cylindrical polar coordinates,
is singular on the axis. 
This leads to rather strong regularity conditions on
tensor fields, as explained in appendix \ref{sec:Conaxi}.
In addition to the axisymmetry, we would like to impose reflection
symmetry about the $z$-axis so that we only need to evolve one half of
the $(r, z)$ plane. This implies that tensor components are either odd
or even functions of $z$.

Let us first deal with one of the regularity conditions for
2-tensors $M_{\alpha\beta}$, which follows from \eqref{eq:introtensorreg}
\begin{equation*}
  \frac{M_{\varphi\varphi}}{r^2 M_{rr}} = 1 + O(r^2) 
\end{equation*}
near the $r$-axis.
For the metric $g_{\alpha\beta}$ this implies
\begin{equation*}
  \label{eq:lambdacond}
  \frac{\lambda^2}{r^2 H_{rr}} = \frac{g_{\varphi\varphi}}{r^2 g_{rr}} 
  = 1 + O(r^2) \, .
\end{equation*}
We enforce this condition by replacing $\lambda$ with a new variable
$\tilde s$ defined by
\begin{equation}
  \label{eq:stildedef}
  \lambda = r \e^{r^2 \tilde s} \sqrt{H_{rr}} \, .
\end{equation}
To satisfy the corresponding condition for the extrinsic curvature, we
introduce $\tilde Y$ via
\begin{equation}
  \label{eq:Ytildedef}
  \K = \frac{\chi_{rr}}{H_{rr}} + r^2 \tilde Y 
\end{equation}
(note that $K_{\varphi\varphi} = \lambda^2 K_\varphi{}^\varphi$).
Similary for the energy-momentum tensor, we set
\begin{equation*}
  \label{tautildedef}
  \tau = \frac{S_{rr}}{H_{rr}} + r^2 \tilde \tau \, .
\end{equation*}
We remark that the definitions of the variables $\tilde s$
\eqref{eq:stildedef} and $\tilde Y$ \eqref{eq:Ytildedef}
can be viewed as a generalization of those in \cite{CHLP} and \cite{GD}.

The remaining dependent variables are redefined by  taking out
systematically the leading order of $r$ and $z$:
\begin{eqnarray*}
  \tilde H_{rr} = H_{rr} \,, \quad 
  \tilde H_{rz} = H_{rz}/(rz) \,, \quad
  \tilde H_{zz} = H_{zz} \,,  \\
  \tilde D_{rrr} = \half \tilde H_{rr,r}/r \,,\quad 
  \tilde D_{rrz} = \half \tilde H_{rz,r}/r \,,\quad
  \tilde D_{rzz} = \half \tilde H_{zz,r}/r \,,\quad \\
  \tilde D_{zrr} = \half \tilde H_{rr,z}/z \,,\quad 
  \tilde D_{zrz} = \half \tilde H_{rz,z}/z \,,\quad
  \tilde D_{zzz} = \half \tilde H_{zz,z}/z \,,\quad \\
  \tilde s_r = \tilde s_{,r}/r \,,\quad 
  \tilde s_z = \tilde s_{,z}/z \,,\quad \\
  \quad \tilde \alpha = \alpha \,,\quad
  \tilde A_r = {\tilde \alpha}^{-1} {\tilde \alpha}_{,r}/r \,,\quad 
  \tilde A_z = {\tilde \alpha}^{-1} {\tilde \alpha}_{,z}/z \,,\quad \\
  \tilde \chi_{rr} = \chi_{rr} \,,\quad
  \tilde \chi_{rz} = \chi_{rz}/(rz) \,,\quad
  \tilde \chi_{zz} = \chi_{zz} \,,\quad \\
  \tilde E^r = E^r/r \,,\quad 
  \tilde E^z = E^z/z \,,\quad 
  \tilde B^\varphi = B^{\varphi}/(rz) \,,\\
  \tilde \theta = \theta \,,\quad
  \tilde Z_r = Z_r /r \,,\quad
  \tilde Z_z = Z_z /z \,,\quad
  \tilde Z^\varphi = Z^\varphi \,,\\
  \tilde \sigma = \sigma \,,\quad
  \tilde \rho_K = \rho_H - \sigma\,,\quad
  \tilde J_r = J_r/r\,,\quad
  \tilde J_z = J_z/z\,,\quad
  \tilde J^\varphi = J^\varphi\,,\\
  \tilde \Sigma_r = \Sigma_r/r\,,\quad
  \tilde \Sigma_z = \Sigma_z/z\,,\quad
  \tilde S_r = S_r/r\,,\quad
  \tilde S_z = S_z/z\,,\\
  \tilde S_{rr} = S_{rr} \,,\quad
  \tilde S_{rz} = S_{rz}/(rz) \,,\quad
  \tilde S_{zz} = S_{zz} \,.
\end{eqnarray*}
  
It can now be verified\footnote{The calculations are rather
  lengthy. We used the computer algebra system REDUCE.}  
that  the Z(2+1)+1 equations can be written in terms of the new
variables $\vec {\tilde u}$ in the form
\begin{equation}
  \label{eq:regconsform}
  \partial_t \vec {\tilde u} 
  + \left[ \alpha \vec {\tilde \F}^{(r^2)} (\vec {\tilde u}) \right]_{,r^2}
  + \left[ \alpha \vec {\tilde \F}^{(z^2)} (\vec {\tilde u}) \right]_{,z^2}  
  = \alpha \vec {\tilde \S} (\vec {\tilde u}) \,,
\end{equation}
where the fluxes $\vec {\tilde \F}$ and the source $\vec {\tilde \S}$
are even in $r$ and $z$ and manifestly regular on the axes
(cf. appendix \ref{sec:regsplit}).

The above transformation of variables does not affect the
hyperbolicity of the system. To see this one should note that 
\begin{eqnarray*}
  \tilde D_{rrr} \simeq D_{rrr}/r \,,\quad 
  \tilde D_{rrz} \simeq D_{rrz}/(r^2 z) \,,\quad
  \tilde D_{rzz} \simeq D_{rzz}/r \,,\\\
  \tilde D_{zrr} \simeq D_{zrr}/z \,,\quad
  \tilde D_{zrz} \simeq D_{zrz}/(r z^2) \,,\quad
  \tilde D_{zzz} \simeq D_{zzz}/z \,,\\
  \tilde s_r \simeq \left( L_r - \frac{D_{rrr}}{H_{rr}} \right)/r^3 \,,\quad 
  \tilde s_z \simeq \left( L_z - \frac{D_{zrr}}{H_{rr}} \right)/(r^2
  z)\,,\\
  \tilde A_r \simeq A_r/r \,,\quad 
  \tilde A_z \simeq A_z/z \,,\\
  \tilde \chi_{rr} \simeq  \chi_{rr} \,,\quad
  \tilde \chi_{rz} \simeq \chi_{rz}/(rz) \,,\quad
  \tilde \chi_{zz} \simeq \chi_{zz} \,,\quad 
  \tilde Y \simeq (\K - \frac{\chi_{rr}}{H_{rr}})/r^2 \,,\\
  \tilde E^r \simeq E^r/r \,,\quad 
  \tilde E^z \simeq E^z/z \,,\quad 
  \tilde B^\varphi \simeq B^{\varphi}/(rz) \,,\\
  \tilde \theta \simeq \theta \,,\quad
  \tilde Z_r \simeq Z_r /r \,,\quad
  \tilde Z_z \simeq Z_z /z \,,\quad
  \tilde Z^\varphi \simeq Z^\varphi \,,\\
  \tilde J_r \simeq J_r/r\,,\quad
  \tilde J_z \simeq J_z/z\,,\\
  \tilde \Sigma_r \simeq \Sigma_r/r\,,\quad
  \tilde \Sigma_z \simeq \Sigma_z/z\,,\quad
  \tilde S_r \simeq S_r/r\,,\quad
  \tilde S_z \simeq S_z/z\,,\\
  \tilde S_{rr} \simeq S_{rr} \,,\quad
  \tilde S_{rz} \simeq S_{rz}/(rz) \,,\quad
  \tilde S_{zz} \simeq S_{zz} \,,
\end{eqnarray*}
where $\simeq$ denotes equality to leading derivative order (the
lower-order terms are absorbed in the source terms)\footnote{Note that
$H_{rr}$ has zero flux and is thus to be treated as a background quantity.}. 
Hence the variables occuring in the fluxes undergo a \emph{linear} 
transformation, which does not affect the hyperbolicity
and leaves the characteristic variables unchanged.

By writing the equations in terms of derivatives with respect to $r^2$
and $z^2$ instead of $r$ and $z$, however, we have ultimately to
multiply the fluxes by $2 r$ and $2 z$, respectively, because
\begin{equation*}
  \partial_{r^2} = \frac{1}{2 r} \partial_r
\end{equation*}
and similarly for $z$. This means that in $(r^2, z^2)$ coordinates, the
characteristic speeds behave like $\sim r$ and $\sim z$, respectively.
For computational purposes it is highly desirable to avoid the
non-uniform characteristic speeds which would occur on a $(r^2, z^2)$
grid.
We therefore recommend working in the original $(r, z)$ grid and
discretizing the derivatives in \eqref{eq:regconsform} with respect to
$r^2$ and $z^2$.
This has another, perhaps more fundamental, advantage: 
in $(r, z)$ coordinates, we
can enforce Neumann boundary conditions for all our variables on the
axes (since they are even functions of $r$ and $z$), whereas in $(r^2, z^2)$
coordinates the boundary conditions on the axes are not known.
The authors of \cite{Nakamura} faced a similar problem but decided to
work on an $(r^2, z^2)$ grid, using extrapolation on the axis, which
led to numerical instabilities.

Choptuik et al \cite{CHLP} and Garfinkle et al \cite {GD} take a
slightly different approach.
They include terms which are formally irregular on axis, but which are
regular once appropriate boundary conditions are imposed.  Following
an idea of Evans \cite{Evans} these terms are differenced with respect
to $r^2$ or $r^3$.

The transformation from conserved to primitive fluid variables and
vice versa can be made manifestly regular on the axes by redefining
the velocities
\begin{equation*}
  \tilde v_r = v_r/r \,, \quad \tilde v_z = v_z/z 
\end{equation*}
and leaving the remaining primitive variables unchanged. Note in
particular that the modified general matter variable
\begin{equation}
  \label{eq:tildetaudef}
  \tilde \tau = r^{-2} \left( \tau - \frac{S_{rr}}{H_{rr}} \right) =
  \rho h W^2 \left( e^{2 r^2 \tilde s} H_{rr} {v^\varphi}^2 
                    - \frac{{\tilde v_r}^2}{H_{rr}} \right)
\end{equation}
is then automatically regular.

To compute the characteristic variables (geometry and matter), one can
start from the regularized conserved variables $\vec {\tilde u}$, compute
the original conserved variables $\vec u$ and evaluate the
characteristic variables (section \ref{sec:Hype} and appendix
\ref{sec:matter_char}). While this transformation
is perfectly regular on the axes, the inverse transformation 
involves negative powers of $r$ and $z$.
In order to implement outer boundary conditions at $r = r_{max}$ ($z =
z_{max}$), one has to ensure that the inverse transformation 
in the $r$-direction ($z$-direction) is regular at 
$z = 0$ ($r = 0$). This can be achieved
by redefining the \emph{characteristic} variables in a similar fashion
as was done for the conserved variables above. We choose to make
the following replacements
\begin{itemize}
  \item for the characteristic variables in the $r$-direction:
  \begin{equation*}
    l_{0,4} \rightarrow \tilde l_{0,4} = z^{-2} \left (l_{0,4} 
         + \frac{z \tilde H_{rz}}{\sqrt{H}} l_{0,5} \right) \,,   
  \end{equation*}  
  and the leading order of $z$ is taken out of the remaining
  characteristic variables

  \item for the characteristic variables in the $z$-direction:
  \begin{eqnarray*}
    l_{0,4} &\rightarrow& \tilde l_{0,4} = r^{-2} \left (l_{0,4} 
         + \frac{r \tilde H_{rz}}{\sqrt{H}} l_{0,5} \right) \,,\\   
    l_{0,6} &\rightarrow& \tilde l_{0,6} = r^{-3} (l_{0,6} - l_{0,5}) \,,\\
    l^\pm_{1,4} &\rightarrow& \tilde l^\pm_{1,4} = 
        r^{-2} (l^\pm_{1,4} - l^\pm_{1,3}) \,,
  \end{eqnarray*}
  and the leading order of $r$ is taken out of the remaining
  characteristic variables
\end{itemize}
We have checked with REDUCE that the transformation from characteristic
variables to conserved variables and {\it vice versa} is then 
manifestly regular at the entire outer boundary.

However, the transformation from characteristic variables in the
$r$-direction ($z$-direction) to conserved variables is still formally 
singular at $r = 0$ ($z=0$), and no regularization
procedure can cure this problem. To understand this, one can note that 
the characteristic variables in the $r$-direction ($z$-direction) 
do not have a definite parity in $r$ ($z$). For this reason, numerical
methods operating in the space of characteristic variables
(typically ones based on the solution of the Riemann problem) 
appear to be unusable near the axes.

%%% Local Variables: 
%%% mode: latex
%%% TeX-master: "master"
%%% End: 

\section{Conclusions}
\label{sec:Concl}

We started out by trying to understand why so many previous attempts
to evolve axisymmetric spacetimes failed because of on-axis
instabilities.
This led to a detailed study of the behaviour of the components of
axisymmetric tensors with respect to cylindrical polar coordinates,
given that the components with respect to cartesian coordinates were
regular in the neighbourhood of the axis.
This suggested, most strongly, that we should make a Kaluza-Klein-like
reduction to a 3-dimensional Lorentzian spacetime \cite{Geroch}, a
step taken earlier by Maeda et al \cite {Maeda}, \cite{Nakamura}.
(A different version of this idea, without rotation, has been pursued
by \cite{CHLP}.)

Choptuik et al's evolution \cite{CHLP} was fully
constrained---elliptic equations were solved for the metric
components, making use of multigrid for computational efficiency.
We made similar choices and obtained similar results.
However in strong field situations, e.g., near-critical collapse of
Brill waves, the diagonal dominance of the matrix describing the
discretized elliptic operators failed, and so the multigrid iterations
failed to converge.

We have not found a satisfactory resolution of this problem, and so we
went to the other extreme---our evolution algorithm involves no
elliptic operators.
To achieve this we have used the Z4 formalism \cite{Bona03},
\cite{Bona04} adapted to our axisymmetric reduction, an evolution
equation for the lapse function and a zero shift gauge.
Our first order hyperbolic system of conservation laws (with sources)
is strongly hyperbolic and, for one choice of parameters, symmetric
hyperbolic. 
In addition the dependent variables have been carefully chosen so that
each and every term in the system is regular on axis.
We have allowed for arbitrary matter sources and have presented
explicitly  the details for a perfect fluid.

Initial conditions have to be imposed via the constraint conditions,
which means solving elliptic equations.
By means of conformal rescalings  on the initial hypersurface one can
avoid the diagonal dominance problem on that hypersurface and so
multigrid techniques are applicable.

One must choose a high resolution shock capturing method for the
evolution of the hyperbolic system.
Our system contains one subtle disadvantage if we choose to work with
numerical algorithms that require a transformation between conserved
and characteristic variables.
The conversion from characteristic variables to our ``regularized''
conserved variables, see section \ref{sec:Reg}, is formally singular on
the axis. 
%Of course there is no problem with using characteristic variables to
%enforce absorbing boundary conditions far from the axis.
However, we have shown that the transformation can be used to impose
outer boundary conditions in a regular way.

Because the evolution does not involve elliptic operators it is
straightforward to introduce adaptive mesh refinement, where
appropriate, to enhance the resolution.

We shall be using our formalism to evolve a number of interesting
axisymmetric spacetimes, including the effects of rotation and perfect
fluid sources.

%%% Local Variables: 
%%% mode: latex
%%% TeX-master: "master"
%%% End: 

\appendix
\newpage
\section{Consequences of Axisymmetry}
\label{sec:Conaxi}

We want to use a $(t, z, r, \varphi)$ chart adapted to the Killing
vector $\xi = \partial/\partial\varphi$.
Unfortunately this chart is singular at the axis $r=0$.
We shall assume \emph{elementary flatness}: in a neighbourhood of the
axis we can introduce local cartesian coordinates $x^A = (x, y)$ such
that
\begin{equation*}
%  \label{eq:polar}
  x = r\cos\varphi, \quad y = r\sin\varphi\quad \Longleftrightarrow
  \quad r = \sqrt{x^2 + y^2}, \quad \varphi = \arctan\frac{y}{x}.
\end{equation*}
With respect to cartesian coordinates the Killing vector is
\begin{equation*}
%  \label{eq:kill1}
  \xi = -y\partial/\partial x + x\partial/\partial y.
\end{equation*}
Notice that this representation is valid everywhere, while 
$\xi = \partial/\partial\varphi$ is valid only for $r>0$.

We say that a scalar function $f(x^A)$ is \emph{regular on axis} if $f$
has a Taylor expansion with respect to $x$ and $y$ about $x^A=0$
convergent in some neighbourhood of $r=0$.
(Throughout this section we are ignoring $t$, $z$ dependencies which
are implicit in all calculations.)
If $f$ is axisymmetric 
\begin{equation}
  \label{eq:kill2}
  \Lie{\xi} f = 0 = \partial f/\partial\varphi\quad\Rightarrow
  \quad f = f(r),\quad r>0.
\end{equation}
If $f$ had a Taylor expansion in $r$ and was regular on axis then the
expansion could contain only even powers of $r$ since
$r=\sqrt{x^2+y^2}$, and $r$ has no such expansion.
While the conclusion is correct, the argument is fallacious because equation
\eqref{eq:kill2} is invalid at $r=0$.
Instead we start from
\begin{equation}
  \label{eq:kill3}
  K\equiv -yf_{,x} + xf_{,y} = 0,
\end{equation}
which is valid everywhere.
In particular we may differentiate \eqref{eq:kill3} an arbitrary
number of times with respect to $x$ and also with respect to $y$.
Then setting $x^A=0$ we obtain  linear recurrence relations between
the Taylor coefficients on axis.
These can be solved to show $f = \sum_n f_n(x^2+y^2)^n$.

Next consider a vector field $u^\alpha$.
For $a = (t,z)$,  $\Lie{\xi}u^a = 0$ implies 
$\partial u^a/\partial\varphi = 0$.
This reduces to the scalar field case and we may deduce 
$u^a = u^a(r^2)$.
For $u^x$ and $u^y$ we have
\begin{equation*}
%  \label{eq:kill4}
  \partial u^x/\partial\varphi + u^y = 0, \quad
  \partial u^y/\partial\varphi - u^x = 0.
\end{equation*}
The general solution for $r>0$ is
\begin{equation*}
  u^x = \widehat{a}(r) \cos\varphi - \widehat{b}(r)\sin\varphi,\quad
  u^y = \widehat{a}(r) \sin\varphi + \widehat{b}(r)\cos\varphi.
\end{equation*}
However in the cartesian chart we have
\begin{equation}
  \label{eq:kill5}
  -yu^x{}_{,x} + x u^x{}_{,y} + u^y = 0, \quad
  -yu^y{}_{,x} + x u^y{}_{,y} - u^x = 0.
\end{equation}
Clearly $u^A = 0$ on axis.  
We write  $\widehat{a} = ra$ etc so that
\begin{equation}
  \label{eq:kill6}
  u^x = xa - yb, \quad u^y = ya + xb.
\end{equation}
We now regard $a$ and $b$ as unknown functions of $x$ and $y$ to be
determined by substituting \eqref{eq:kill6} into \eqref{eq:kill5},
differentiating the latter an arbitrary number of times, and then
solving the recurrence relations for the Taylor coefficients of $a$
and $b$.
Again we find that $a$ and $b$ are even functions of $r$.
Thus in the $(t, z, r, \varphi)$ chart an axisymmetric vector field
which is regular on axis must take the form
\begin{equation}
  \label{eq:kill7}
  u^\alpha = (A, B, rC, D),
\end{equation}
where $A$, $B$, $C$ and $D$ are functions of $t$, $z$ and $r^2$.

Next consider an axisymmetric covector field $\omega_\alpha$
which is regular on axis.
For $a = (t,z)$,  $\Lie{\xi}\omega_a = 0$ implies 
$\partial \omega_a/\partial\varphi = 0$.
This reduces to the scalar field case and we may deduce 
$\omega_a = \omega_a(r^2)$.
For the other indices we find
\begin{equation}
  \label{eq:kill10}
  -y\omega_{x,x} + x\omega_{x,y} + \omega_y = 0, \quad
  -y\omega_{y,x} + x\omega_{y,y} - \omega_x = 0,
\end{equation}
which is equivalent to \eqref{eq:kill5}, interchanging $u^A$ and
$\omega_A$.
We therefore deduce the analogue of \eqref{eq:kill6} and hence, in
polar coordinates,
\begin{equation}
  \label{eq:kill8}
  \omega_\alpha = (E, F, rG, r^2H),
\end{equation}
where $E$, $F$, $G$ and $H$ are functions of $t$, $z$ and $r^2$.

Finally we consider a symmetric valence 2 axisymmetric tensor field
$M_{\alpha\beta}$ which is both axisymmetric and regular on axis.
For $(a, b) = (t, z)$ we have $\Lie{\xi}M_{ab}=0$ and so 
$M_{ab} = M_{ab}(r^2)$.
Letting $a = (t, z)$ we have
\begin{equation}
  \label{eq:kill11}
  -yM_{ax,x} + xM_{ax,y} + M_{ay} = 0, \quad
  -yM_{ay,x} + xM_{ay,y} - M_{ax} = 0.
\end{equation}
This is essentially the same as \eqref{eq:kill10} and we may deduce
$M_{ar} = rA_a(r^2)$ and $M_{a\varphi} = r^2B_a(r^2)$.
The remaining Killing equations are
\begin{equation*}
  -yM_{xx,x} + xM_{xx,y} + 2M_{xy} = 0,\quad
  -yM_{yy,x} + xM_{yy,y} - 2M_{xy} = 0,\quad
  -yM_{xy,x} + xM_{xy,y} + M_{yy} - M_{xx} = 0.
\end{equation*}
If we introduce new variables
$u = \half(M_{xx} + M_{yy})$, $v = \half(M_{xx} - M_{yy})$ and 
$w=M_{xy}$ then
\begin{equation*}
  -yu_{,x} + xu_{,y} = 0
\end{equation*}
which implies $u = u(r^2)$.
The remaining equations are
\begin{equation}
  \label{eq:kill12}
  -yv_{,x} + xv_{,y} + 2w=0, \quad
  -yw_{,x} + xw_{,y} - 2v=0.
\end{equation}
For $r>0$ these can be written as
\begin{equation*}
  v_{,\varphi} + 2w=0, \quad w_{,\varphi} - 2w=0,
\end{equation*}
so that
\begin{equation*}
  v = \widehat a(r)\cos2\varphi - \widehat b(r)\sin2\varphi,\quad
  w = \widehat a(r)\sin2\varphi + \widehat b(r)\cos2\varphi,
\end{equation*}
where $\widehat a$ and $\widehat b$ are arbitrary functions of $r$.
If we set $a = \widehat a/r^2$ and $b = \widehat b/r^2$ then
\begin{equation}
  \label{eq:kill13}
  v = (x^2-y^2)a - 2xyb, \quad w = 2xya + (x^2-y^2)b.
\end{equation}
(Note that \eqref{eq:kill12} and its first derivatives imply that
$v$, $w$ and their first derivatives vanish on axis, consistent with
\eqref{eq:kill13}.) 
Substituting \eqref{eq:kill13} into \eqref{eq:kill12} gives
\begin{eqnarray}
  \label{eq:kill14}
  x^3a_{,y} - x^2y(a_{,x} + 2b_{,y}) - 
  xy^2(a_{,y} - 2b_{,x}) + y^3a_{,x} =& 0,\\
  x^3b_{,y} + x^2y(-b_{,x} + 2a_{,y}) + 
  xy^2(-2a_{,x} - b_{,y}) + y^3b_{,x} =& 0.
\end{eqnarray}
Differentiating these many times and, proceeding as in the scalar and
vector cases we conclude  that $a$ and $b$ are functions of $r^2$.
Thus
\begin{equation*}
  M_{xx} = u + (x^2-y^2)a -2xyb, \quad
  M_{yy} = u - (x^2-y^2)a + 2xyb,\quad
  M_{xy} = 2xya + (x^2-y^2)b.
\end{equation*}
Re-expressing these as polar components we obtain
\begin{equation}
  \label{eq:kill15}
  M_{rr} = u + r^2a, \quad M_{r\varphi} = r^3b, \quad
  M_{\varphi\varphi} = r^2(u - r^2a).
\end{equation}
Finally combining all of the results we have
\begin{equation}
  \label{eq:kill9}
  M_{\alpha\beta} = \left( \begin{array}{cccc} A & B & r D & r^2 F \\
      B & C & r E & r^2 G \\ r D & r E & H + r^2 J & r^3 K \\ 
      r^2 F & r^2 G & r^3 K & r^2 \left( H - r^2 J \right) \end{array} \right)
  ,
\end{equation}
where $A,B,\ldots,K$ are functions of $t$, $z$ and $r^2$.

%%% Local Variables: 
%%% mode: latex
%%% TeX-master: "master"
%%% End: 

\newpage
\section{Transformation from characteristic to conserved variables}
\label{sec:InvTrafo}

The conserved variables are given in terms of the characteristic variables by

\begin{eqnarray*}
  D_{\n\n\n} &=& -\frac{1}{f} l_{0,1} + \left( \frac{f(m-2)}{2(f-1)} + 1
    \right) (l_{1,1}^+ - l_{1,1}^-) - \half (l_{1,3}^+ - l_{1,3}^-
    + l_{1,4}^+ - l_{1,4}^-) \\&&  
    + \frac{1}{2f} \left( l_f^+ + l_f^- \right) \,,\\
  D_{\n\n\p} &=& -\frac{1}{fm} l_{0,2} + \frac{(m-2)}{2m} (l_{0,3} +
    l_{0,5} + l_{0,6})
    - \frac{(fm-2)}{2fm} l_{0,7} + \half (l_{1,2}^+ - l_{1,2}^-) \,,\\
  D_{\n\p\p} &=& \half (l_{1,3}^+ - l_{1,3}^-) \,,\\
  D_{\p\n\n} &=& l_{0,3} \,,\\
  D_{\p\p\n} &=& l_{0,4} \,,\\
  D_{\p\p\p} &=& l_{0,5} \,,\\
  L_\n &=& \half (l_{1,4}^+ - l_{1,4}^-) \,,\\
  L_\p &=& l_{0,6} \,,\\
  A_\n &=& \frac{f(m-2)}{2(f-1)} (l_{1,1}^+ - l_{1,1}^-) 
    + \half \left( l_f^+ + l_f^- \right) \,,\\
  A_\p &=& l_{0,7} \,,\\
  \chi_{\n\n} &=& \left( \frac{f(m-2)}{2(f-1)} + 1 \right) 
    (l_{1,1}^+ + l_{1,1}^-) - \half (l_{1,3}^+ + l_{1,3}^- 
     + l_{1,4}^+ + l_{1,4}^-) \\&&
    + \frac{1}{2\sqrt{f}} \left( l_f^+ - l_f^- \right) \,,\\
  \chi_{\n\p} &=& \half (l_{1,2}^+ + l_{1,2}^-) \,,\\
  \chi_{\p\p} &=& \half (l_{1,3}^+ + l_{1,3}^-) \,,\\
  \K &=& \half (l_{1,4}^+ + l_{1,4}^-) \,,\\
  E^\n &=& \half (l_{1,5}^+ + l_{1,5}^-) \,,\\
  E^\p &=& \half (l_{1,6}^+ + l_{1,6}^-) \,,\\
  B^\varphi &=& -\half (l_{1,6}^+ - l_{1,6}^-) \,,\\
  \theta &=& \half (l_{1,1}^+ + l_{1,1}^-) \,,\\
  Z_\n &=& - l_{0,4} + \half (-l_{1,1}^+ + l_{1,1}^- + l_{1,3}^+ - l_{1,3}^-
    + l_{1,4}^+ - l_{1,4}^-) \,,\\
  Z_\p &=& \half (l_{0,3} - l_{0,5} + l_{0,6} + l_{0,7}) 
    - \half(l_{1,2}^+ - l_{1,2}^-) \,,\\
  Z^\varphi &=& -\frac{1}{4} (l_{1,5}^+ - l_{1,5}^-) \,.  
\end{eqnarray*}

Tensor components can then be computed as, for instance,
\begin{equation*}
  \chi_{AB} = \mu_A \mu_B \chi_{\n\n} + 2 \mu_{(A} \pi_{B)} \chi_{\n\p} 
    + \pi_A \pi_B \chi_{\p\p} \, .
\end{equation*}

%%% Local Variables: 
%%% mode: latex
%%% TeX-master: "master"
%%% End: 

\section{Regular conservation form}
\label{sec:regsplit}

In this appendix we provide some details of the regular form of the
Z(2+1)+1 equations outlined in section \ref{sec:Reg}.
The equations are written in conservation form \eqref{eq:regconsform}
\begin{equation*}
  \partial_t \vec {\tilde u} 
  + \left[ \alpha \vec {\tilde \F}^{(r^2)} (\vec {\tilde u}) \right]_{,r^2}
  + \left[ \alpha \vec {\tilde \F}^{(z^2)} (\vec {\tilde u}) \right]_{,z^2}  
  = \alpha \vec {\tilde \S} (\vec {\tilde u}) \,,
\end{equation*}
where the modified variables $\vec {\tilde u}$ are defined in
section \ref{sec:Reg}.

The fluxes $\vec {\tilde \F}^{(r^2)}, \, \vec {\tilde \F}^{(z^2)}$ 
and sources $\vec {\tilde \S}$ have been obtained with the
computer algebra system REDUCE and have been typeset
automatically using the TeX REDUCE Interface (TRI).
From the same source we generated C++ code using the REDUCE source
code optimization packages SCOPE and GENTRAN.
The resulting expressions are rather lengthy.
%and are given in full in the electronic version of this paper.
To get some idea of what is typically involved we state here the
fluxes and source for the variable $\tilde Y$.

Let us introduce the shorthand
\begin{equation*}
  H = \tilde H_{rr} \tilde H_{zz} - r^2 z^2 \tilde H_{rz}^2
\end{equation*}
for the determinant of the 2-metric.
The fluxes are given by

\begin{eqnarray*}
\tilde \F^{(r^2)}_{\, \tilde Y} &=& 
  2H^{-1}\tilde D_{rrr} r^{2}z^{2}\tilde H_{rr} ^{-2}\tilde H_{rz} ^{2}
  -2H^{-1}\tilde D_{rzz} 
  -4H^{-1}\tilde D_{zrr} z^{2}\tilde H_{rr} ^{-1}\tilde H_{rz} 
  +4H^{-1}\tilde D_{zrz} z^{2}\\ &&
  +2\tilde H_{rr} ^{-1} (-\tilde A_r +2\tilde Z_r)
  +2H^{-1}r^{2}z^{2}\tilde H_{rz} 
  (r^{2}\tilde H_{rr} ^{-1}\tilde H_{rz} \tilde s_r -\tilde s_z) \,,\\ 
  \tilde \F^{(z^2)}_{\, \tilde Y} &=& 
  2H^{-1}z^{2} (-r^{2}\tilde H_{rz} \tilde s_r +\tilde H_{rr} \tilde s_z)\,,
\end{eqnarray*}
and the source term is
\begin{eqnarray*}
\tilde \S_{\, \tilde Y} &=&  
  -\kappa \tilde \tau 
  +2H^{-1}\tilde \chi_{rr} ^{2}z^{2}\tilde H_{rr} ^{
  -2}\tilde H_{rz} ^{2}
  +2\tilde \chi_{rr} \tilde H_{rr} ^{
  -1}\tilde Y 
  +H^{-1}\tilde \chi_{rr} z^{2}\tilde H_{rr} ^{
  -1}\tilde H_{rz} 
  (r^{2}\tilde H_{rz} \tilde Y 
    -4\tilde \chi_{rz} 
  )
  \\ &&
  +H^{-1}\tilde \chi_{zz} \tilde H_{rr} \tilde Y 
  -4H^{-1}\tilde D_{rrr} ^{2}r^{2}z^{2}\tilde H_{rr} ^{
  -3}\tilde H_{rz} ^{2}\\ &&
  -H^{-2}\tilde D_{rrr} ^{2}r^{4}z^{4}\tilde H_{rr} ^{
  -3}\tilde H_{rz} ^{4}
  +4H^{-1}\tilde D_{rrr} \tilde D_{rrz} r^{2}z^{2}
  \tilde H_{rr} ^{-2}\tilde H_{rz} \\ &&
  +2H^{-2}\tilde D_{rrr} \tilde D_{rrz} r^{4}z^{4}
  \tilde H_{rr} ^{-2}\tilde H_{rz} ^{3}
  +4H^{-1}\tilde D_{rrr} \tilde D_{zrr} z^{2}\tilde H_{rr} 
  ^{
  -2}\tilde H_{rz} \\ &&
  +2H^{-2}\tilde D_{rrr} \tilde D_{zrr} r^{2}z^{4}
  \tilde H_{rr} ^{-2}\tilde H_{rz} ^{3}
  -2H^{-2}\tilde D_{rrr} \tilde D_{zzz} z^{2}\tilde H_{rz} 
  \\ &&
  +H^{-1}\tilde D_{rrr} \tilde H_{rr} ^{
  -2}\tilde H_{rz} 
  (-5r^{4}z^{2}\tilde H_{rz} \tilde s_r 
    -r^{2}z^{2}\tilde A_r \tilde H_{rz} 
    +4r^{2}z^{2}\tilde H_{rr} \tilde s_z 
    -2r^{2}z^{2}\tilde H_{rz} \tilde s 
    \\ && \qquad \qquad \qquad \qquad \quad
    +4r^{2}z^{2}\tilde H_{rz} \tilde Z_r 
    +2z^{2}\tilde A_z \tilde H_{rr} 
    -4z^{2}\tilde H_{rr} \tilde Z_z 
    +2\tilde H_{rr} 
  )\\ &&
  +H^{-2}\tilde D_{rrr} r^{2}z^{2}\tilde H_{rr} ^{
  -2}\tilde H_{rz} ^{3}
  (-r^{4}z^{2}\tilde H_{rz} \tilde s_r 
    +r^{2}z^{2}\tilde H_{rr} \tilde s_z 
    +2r^{2}z^{2}\tilde H_{rz} \tilde s 
    +2z^{2}\tilde H_{rz} 
    +2\tilde H_{rr} 
  ) \\ &&
  -2H^{-2}\tilde D_{rrz} \tilde D_{rzz} r^{2}z^{2}
  \tilde H_{rz} 
  -4H^{-1}\tilde D_{rrz} \tilde D_{zrr} z^{2}\tilde H_{rr} 
  ^{
  -1}
  -6H^{-2}\tilde D_{rrz} \tilde D_{zrr} r^{2}z^{4}
  \tilde H_{rr} ^{-1}\tilde H_{rz} ^{2} \\ &&
  +4H^{-2}\tilde D_{rrz} \tilde D_{zrz} r^{2}z^{4}
  \tilde H_{rz} 
  +2H^{-2}\tilde D_{rrz} \tilde D_{zzz} z^{2}\tilde H_{rr} 
  \\ &&
  +2H^{-1}\tilde D_{rrz} 
  (2r^{4}z^{2}\tilde H_{rr} ^{
    -1}\tilde H_{rz} \tilde s_r 
    +r^{2}z^{2}\tilde A_r \tilde H_{rr} ^{
    -1}\tilde H_{rz} 
    -2r^{2}z^{2}\tilde H_{rr} ^{
    -1}\tilde H_{rz} \tilde Z_r \\ && \qquad \qquad \qquad
    -r^{2}z^{2}\tilde s_z 
    -z^{2}\tilde A_z 
    +2z^{2}\tilde Z_z 
    -1
  ) \\ &&
  +2H^{-2}\tilde D_{rrz} r^{2}z^{2}\tilde H_{rz} ^{2}
  (r^{4}z^{2}\tilde H_{rr} ^{-1}\tilde H_{rz} \tilde s_r 
    -2r^{2}z^{2}\tilde H_{rr} ^{
    -1}\tilde H_{rz} \tilde s 
    -r^{2}z^{2}\tilde s_z 
    -z^{2}\tilde H_{rr} ^{-1}\tilde H_{rz} 
    -1
  ) \\ &&
  +H^{-2}\tilde D_{rzz} ^{2}\tilde H_{rr} 
  +4H^{-2}\tilde D_{rzz} \tilde D_{zrr} z^{2}\tilde H_{rz} 
  -4H^{-2}\tilde D_{rzz} \tilde D_{zrz} z^{2}\tilde H_{rr} 
  \\ &&
  +H^{-1}\tilde D_{rzz} 
  (-r^{2}\tilde s_r 
    -\tilde A_r 
    -2\tilde s 
  )
  +H^{-2}\tilde D_{rzz} r^{2}z^{2}\tilde H_{rz} 
  (-r^{2}\tilde H_{rz} \tilde s_r 
    +\tilde H_{rr} \tilde s_z 
    +2\tilde H_{rz} \tilde s 
  ) \\ &&
  +2H^{-1}\tilde D_{zrr} z^{2}
  (r^{2}\tilde H_{rr} ^{-1}\tilde H_{rz} \tilde s_r 
    -\tilde A_r \tilde H_{rr} ^{-1}\tilde H_{rz} 
    -\tilde s_z 
  ) \\ &&
  +H^{-2}\tilde D_{zrr} z^{4}\tilde H_{rz} ^{2}
  (r^{4}\tilde H_{rr} ^{-1}\tilde H_{rz} \tilde s_r 
    -2r^{2}\tilde H_{rr} ^{-1}\tilde H_{rz} \tilde s 
    -r^{2}\tilde s_z 
    -4\tilde H_{rr} ^{-1}\tilde H_{rz} 
  ) \\ &&
  +2H^{-1}\tilde D_{zrz} z^{2}
  (\tilde A_r 
    +2\tilde s 
  )
  +2H^{-2}\tilde D_{zrz} z^{4}\tilde H_{rz} 
  (-r^{4}\tilde H_{rz} \tilde s_r 
    +r^{2}\tilde H_{rr} \tilde s_z 
    +2r^{2}\tilde H_{rz} \tilde s 
    +2\tilde H_{rz} 
  ) \\ &&
  +H^{-2}\tilde D_{zzz} z^{2}\tilde H_{rr} 
  (r^{2}\tilde H_{rz} \tilde s_r 
    -\tilde H_{rr} \tilde s_z 
    -2\tilde H_{rz} \tilde s 
  )
  -
  \frac{1}{
        2}e^{2 r^2 \tilde s} Hz^{2}\tilde {E^z} ^{2} \\ &&
  +e^{2 r^2 \tilde s} r^{2}
  (-z^{4}\tilde {E^z} ^{2}\tilde H_{rz} ^{2}
    +z^{2}\tilde {B^\varphi} ^{2}\tilde H_{rr} 
    -2z^{2}\tilde {E^r} \tilde {E^z} \tilde H_{rr} 
    \tilde H_{rz} 
    -\tilde {E^r} ^{2}\tilde H_{rr} ^{2}
  )
  -r^{2}\tilde A_r \tilde H_{rr} ^{
  -1}\tilde s_r \\ &&
  +2r^{2}\tilde H_{rr} ^{-1}\tilde s_r \tilde Z_r 
  +r^{2}\tilde Y ^{2}
  -2\tilde A_r \tilde H_{rr} ^{-1}\tilde s 
  +2\tilde A_r \tilde H_{rr} ^{-1}\tilde Z_r 
  +4\tilde H_{rr} ^{-1}\tilde s \tilde Z_r 
  -2\tilde \theta \tilde Y \\ &&
  +H^{-1}
  (-r^{6}z^{2}\tilde H_{rr} ^{-1}\tilde H_{rz} ^{2}
    \tilde s_r ^{2}
    -4r^{4}z^{2}\tilde H_{rr} ^{
    -1}\tilde H_{rz} ^{2}\tilde s \tilde s_r 
    +2r^{4}z^{2}\tilde H_{rr} ^{
    -1}\tilde H_{rz} ^{2}\tilde s_r \tilde Z_r \\ && \qquad \qquad
    +2r^{4}z^{2}\tilde H_{rz} \tilde s_r \tilde s_z 
    -2r^{2}z^{2}\tilde A_r \tilde H_{rr} ^{
    -1}\tilde H_{rz} ^{2}\tilde s 
    -2r^{2}z^{2}\tilde \chi_{rz} \tilde H_{rz} \tilde Y 
    -r^{2}z^{2}\tilde H_{rr} \tilde s_z ^{2} \\ &&\qquad \qquad
    -4r^{2}z^{2}\tilde H_{rr} ^{
    -1}\tilde H_{rz} ^{2}\tilde s ^{2}
    +4r^{2}z^{2}\tilde H_{rr} ^{
    -1}\tilde H_{rz} ^{2}\tilde s \tilde Z_r 
    -3r^{2}z^{2}\tilde H_{rr} ^{
    -1}\tilde H_{rz} ^{2}\tilde s_r \\ &&\qquad \qquad
    +4r^{2}z^{2}\tilde H_{rz} \tilde s \tilde s_z 
    -2r^{2}z^{2}\tilde H_{rz} \tilde s_r \tilde Z_z 
    -2r^{2}z^{2}\tilde H_{rz} \tilde s_z \tilde Z_r 
    -r^{2}\tilde H_{rz} \tilde s_r \\ &&\qquad \qquad
    +2z^{2}\tilde A_z \tilde H_{rz} \tilde s 
    +2z^{2}\tilde \chi_{rz} ^{2}
    +2z^{2}\tilde H_{rr} \tilde s_z \tilde Z_z 
    -6z^{2}\tilde H_{rr} ^{-1}\tilde H_{rz} ^{2}\tilde s \\ &&\qquad \qquad
    -4z^{2}\tilde H_{rz} \tilde s \tilde Z_z 
    +4z^{2}\tilde H_{rz} \tilde s_z 
    +\tilde H_{rr} \tilde s_z 
    +2\tilde H_{rz} \tilde s 
  ) \\ &&
  +H^{-2}z^{2}\tilde H_{rz} ^{2}
  (r^{4}z^{2}\tilde H_{rr} ^{-1}\tilde H_{rz} ^{2}
    \tilde s_r 
    -r^{4}\tilde H_{rz} \tilde s_r 
    -2r^{2}z^{2}\tilde H_{rr} ^{
    -1}\tilde H_{rz} ^{2}\tilde s 
    -r^{2}z^{2}\tilde H_{rz} \tilde s_z \\ && \qquad \qquad \qquad
    +r^{2}\tilde H_{rr} \tilde s_z 
    +2r^{2}\tilde H_{rz} \tilde s 
    -z^{2}\tilde H_{rr} ^{-1}\tilde H_{rz} ^{2}
  ) \, .
\end{eqnarray*}

Note that the expressions are manifestly regular on the axes and even
in $r$ and $z$.

%%% Local Variables: 
%%% mode: latex
%%% TeX-master: "master"
%%% End: 

\section{Matter evolution equations}
\label{sec:matter}

\subsection{Conservation form}
\label{sec:matter_cons}

The matter evolution equations (\ref{eq:d0rhoH}--\ref{eq:d0sigma}) can
clearly be written in conservation form
\begin{equation*}
%  \label{eq:matter_consform}  
  \partial_t \vec u + \left[ \alpha \vec \F^D (\vec u) \right]_{,D} 
  = \alpha \vec \S (\vec u) \,.
\end{equation*}
Following \cite{Banyuls}, we replace $\rho_H$ with  $\rho_K = \rho_H - \sigma$ 
(kinetic energy) and regard as the set of conserved variables
\begin{equation}
  \label{eq:matterconsvars}
  \vec u = (\rho_K, J_A, J^\varphi, \sigma)^T \,.
\end{equation}
The fluxes are
\begin{eqnarray*}
  \F^D{}_{\rho_K} &=& J^D - \Sigma^D \,,\\
  \F^D{}_{J_A} &=& S_A{}^D \,,\\
  \F^D{}_{J^\varphi} &=& S^D \,,\\
  \F^D{}_{\sigma} &=& \Sigma^D \,,
\end{eqnarray*}
and the source terms are
\begin{eqnarray*}
  \S_{\rho_K} &=& (\Sigma^A - J^A)(D^I{}_A + L_A)  
    + \K (\tau - \sigma) + \chi_{AB} S^{AB} - J^A A_A 
    + \chi \rho_K \nonumber\\ && + \lambda^2 E^A S_A \,,\\
  \S_{J_A} &=&  - S_{AB} (A^B + L^B) + J_A (\chi + \K) 
    - A_A \rho_H + L_A \tau \nonumber\\&&
    + \lambda^2 (E_A J^\varphi + \epsilon_{AB} S^B B^\varphi) \,,\\
  \S_{J^\varphi} &=& - (D^I{}_A + 3 L_A) S^A + J^\varphi (\chi + 3 \K)
    \,,\\
  \S_{\sigma} &=& - (D^I{}_A + L_A) \Sigma^A + \sigma (\chi + \K) \,. 
\end{eqnarray*}

\subsection{Perfect fluid}
\label{sec:matter_fluid}

To evaluate the characteristic structure, we need to specify the matter model.
Here, we consider a perfect fluid with four-velocity $u^\alpha$,
normalized such that 
\begin{equation*}
  u_\alpha u^\alpha = -1 \,,
\end{equation*}
rest mass density $\rho$, pressure $p$ and internal energy $\epsilon$. 
The dependence of the pressure on the density and the internal energy is
given by the equation of state 
\begin{equation}
  \label{eq:EOS}
  p = p(\rho, \epsilon) \,.
\end{equation}
With those definitions, the number density is
\begin{equation*}
  N^\alpha = \rho u^\alpha
\end{equation*}
and the energy-momentum tensor is given by
\begin{equation*}
  T^{\alpha\beta} = \rho h u^\alpha u^\beta + p g^{\alpha\beta} \, ,
\end{equation*}
where $h$ is the specific enthalpy,
\begin{equation}
  \label{eq:enthalpy}
  h = 1 + \epsilon + \frac{p}{\rho} \, .
\end{equation}
The Lorentz factor is defined as
\begin{equation}
  \label{eq:Lorentz}
  W \equiv - u^\alpha n_\alpha \, .
\end{equation}
Observers who are at rest in a slice $\Sigma(t)$ (i.e., who have
four-velocity $n^\alpha$) measure a coordinate velocity
\begin{equation*}
  v^A = W^{-1}  h_\alpha{}^A u^\alpha \, ,
\end{equation*}
and the angular velocity is
\begin{equation*}
  v^\varphi =  W^{-1} \lambda^{-2} \xi_\alpha u^\alpha \,.
\end{equation*}
Hence we obtain the familiar relation
\begin{equation}
  \label{eq:Lorentzrelation}
  W = (1 - v^2)^{-1/2} \,,
\end{equation}
where
\begin{equation*}
  v^2 =  v_A v^A + \lambda^2 {v^\varphi}^2 \,.
\end{equation*}

The variables
\begin{equation*}
  \vec w = (v_A, v^\varphi, \rho, p, \epsilon, h, W)^T
\end{equation*}
are called \emph{primitive variables}.\footnote{Note only five of
  those are independent because of \eqref{eq:EOS}, \eqref{eq:enthalpy}
  and \eqref{eq:Lorentzrelation}.}
The conserved variables can be expressed in terms of the primitive 
variables as
\begin{eqnarray}
  \label{eq:prim2cons}
  \rho_K &=& \rho h W^2 - p - \rho W \,,\nonumber\\
  J_A &=& \rho h W^2 v_A \,,\nonumber\\
  J^\varphi &=& \rho h W^2 v^\varphi \,,\\
  \sigma &=& \rho W \,,\nonumber
\end{eqnarray}
and the remaining matter variables are
\begin{eqnarray}
  \label{eq:prim2noncons}
  \tau &=& \rho h W^2 \lambda^2 {v^\varphi}^2 + p \,,\nonumber\\
  S_A &=& \rho h W^2 v^\varphi v_A \,,\nonumber\\
  S_{AB} &=& \rho h W^2 v_A v_B + p H_{AB} \,,\\
  \Sigma_A &=& \rho W v_A \,.\nonumber
\end{eqnarray}

\subsection{Characteristic decomposition}
\label{sec:matter_char}

The characteristic decomposition for 3+1 general relativistic hydrodynamics
was first derived by the Valencia group \cite{Banyuls}. The application to our
(2+1)+1 system is straightforward.
%\footnote{Note however the additional source terms in our system.}
Our method differs slightly in that we choose a general orthonormal basis 
$(\mu^A, \pi^A)$ in two-space as in section \ref{sec:Hype}
and project vectors along $\mu$ (index $\parallel$) and $\pi$ (index $\perp$). 
Following the notation of \cite{Font}, we introduce a few abbreviations. 
From the equation of state \eqref{eq:EOS}, we form
\begin{equation*}
  \chi \equiv \frac{\partial p}{\partial \rho} \, , \qquad \kappa \equiv
  \frac{\partial p}{\partial \epsilon} \, , \qquad h c_s^2 \equiv \chi +
  \frac{p}{\rho^2} \kappa \, , 
\end{equation*}
where $c_s$ is known as the \emph{sound speed}. 
Also set\footnote{Our definitions of $\xi$
  and $\Delta$ differ from those in \cite{Font} by a factor of
  $\lambda^2$ to ensure regularity (see section \ref{sec:Reg}).
  We have defined $\Kf^{-1}$ instead of $\Kf$ to allow for the special
  case of the ultrarelativistic equation of state, for which $\Kf^{-1} = 0$.
  As a consequence, $\Delta$ above has been multiplied by $\Kf^{-1}$ and the 
  characteristic variable $l_{0,1}$ has been divided by $\Kf^{-1}$.
}
\begin{eqnarray*}
  \Kf^{-1} = 1 - \frac{c_s^2 \rho}{\kappa} \,, \qquad
  \V^\pm = \frac{v_\n - \lambda_s^\pm}{1 - v_\n \lambda_s^\pm} 
    \,, \qquad
  \A^\pm = \frac{1 - v_\n^2}{1 - v_\n \lambda_s^\pm} \,, 
  \nonumber\\
  \C^\pm = v_\n - \V^\pm \,, \qquad
  \xi = 1 - v_\n^2 \,, \qquad
  \Delta = h^3 W (1 - \Kf^{-1})(\C^+ - \C^-) \xi \,.
\end{eqnarray*}

The system is found to be strongly hyperbolic.
The characteristic speeds in the $\mu$-direction are 
\begin{eqnarray*}
  \lambda_0 &=& v_\n \, , \nonumber\\
  \lambda_s^\pm &=& \frac{1}{1 - v^2 c_s^2 } \left\{ v_\n (1-c_s^2) 
     \pm c_s \sqrt{ (1 - v^2)
     \left[ (1 - v^2 c_s^2) - v_\n^2 (1 - c_s^2) \right] } \right\} \, .
\end{eqnarray*}
The characteristic variables (corresponding to the left eigenvectors) are
\begin{eqnarray*}
  l_{0,1} &=& \frac{W}{1 - \Kf^{-1}} \left\{ h \sigma - W (\sigma + \rho_K)
    + W (v_\n J_\n + v_\p J_\p + \lambda^2 v^\varphi J^\varphi) \right\} \,,\\
  l_{0,2} &=& \frac{1}{h \xi} \left\{ - v_\p (\sigma + \rho_K) 
    + v_\n v_\p J_\n + (1 - v_\n^2) J_\p \right\} \,, \\
  l_{0,3} &=& \frac{1}{h \xi} \left\{ - v^\varphi (\sigma + \rho_K)
    + v^\varphi v_\n J_\n + (1 - v_\n^2) J^\varphi \right\} \,,\\
  l_s^\mp &=& \frac{h^2}{\Delta} \left\{ \Kf^{-1} h W \V^\pm \xi \sigma
    + \left[ \Kf^{-1} - \A^\pm - (2 - \Kf^{-1}) v_\n \right] J_\n 
    \right. \nonumber\\ && \qquad
    + (2 - \Kf^{-1}) \V^\pm W^2 \xi (v_\n J_\n + v_\p J_\p 
    + \lambda^2 v^\varphi J^\varphi) \nonumber\\ && \qquad \left.
    + \left[ (\Kf^{-1} - 1) \left( - v_\n + \V^\pm (W^2 \xi - 1) \right) 
    - W^2 \V^\pm \xi \right] (\sigma + \rho_K)
   \right\} \,.
\end{eqnarray*}
The inverse transformation (corresponding to the right eigenvectors)
is given by
\begin{eqnarray*}
  \sigma &=& \frac{1}{h W} l_{0,1} + W(v_\p l_{0,2} + \lambda^2
    v^\varphi l_{0,3}) + l_s^+ + l_s^- \,,\\
  J_\n &=& \Kf^{-1} v_\n l_{0,1} + 2 h W^2 v_\n (v_\p l_{0,2} + \lambda^2
    v^\varphi l_{0,3}) + h W (\C^+ l_s^+ + \C^- l_s^- ) \,,\\
  J_\p &=& \Kf^{-1} v_\p l_{0,1} + h l_{0,2} + 2 h W^2 v_\p (v_\p l_{0,2} +
    \lambda^2 v^\varphi l_{0,3}) + h W v_\p (l_s^+ + l_s^-) \,,\\
  J^\varphi &=& \Kf^{-1} v^\varphi l_{0,1} + h l_{0,3} + 2 h W^2 v^\varphi 
    (v_\p l_{0,2} + \lambda^2 v^\varphi l_{0,3}) + h W v^\varphi (l_s^+ +
    l_s^-) \,,\\
  \rho_K &=& \left( \Kf^{-1} - \frac{1}{h W} \right) l_{0,1} + W
    (2 h W - 1) (v_\p l_{0,2} + \lambda^2 v^\varphi l_{0,3}) \nonumber\\&&
    + h W (\A^+ l_s^+ + \A^- l_s^-) - l_s^+ - l_s^- \,.
\end{eqnarray*}

\subsection{Transformation from conserved to primitive variables}
\label{sec:matter_trafo}

The conserved matter variables \eqref{eq:matterconsvars} are the ones
that are evolved in a numerical algorithm. To compute the remaining
matter variables \eqref{eq:prim2noncons} and the eigenvectors, 
the primitive variables have to be calculated from the conserved
variables as an intermediate step.
This transformation is much more involved than the opposite
direction \eqref{eq:prim2cons}.
To make it explicit, we have to specify an equation of state. 
Here, we consider the ultrarelativistic equation of state,
\begin{equation}
  \label{eq:EOS_spec}
  p = (\Gamma - 1) \rho_{tot} = (\Gamma - 1) \rho (\epsilon + 1)
    = \frac{\Gamma - 1}{\Gamma} \rho h \,,
\end{equation}
where $\rho_{tot}$ is the total energy density.

Suppose we are given the conserved variables, and also form $\rho_H =
\rho_K + \sigma$. Consider the quantity
\begin{equation*}
  J^2 \equiv J_A J^A + \lambda^2 {J^\varphi}^2 \, .
\end{equation*}
Using \eqref{eq:prim2cons}, \eqref{eq:EOS_spec} and
\eqref{eq:Lorentzrelation},
we can express $J^2$ and $\rho_H$ in terms of the primitive variables as
\begin{eqnarray*}
  J^2 &=& \left( \frac{\Gamma}{\Gamma-1} \right)^2 p^2 W^2 (W^2 -1)
    \, , \nonumber\\
  \rho_H &=& p \, \left(\frac{\Gamma}{\Gamma-1} W^2 - 1\right) \, .
\end{eqnarray*}
Eliminating $W$ yields an equation for the pressure in terms of
conserved variables:
\begin{equation*}
  p = - 2 \beta \rho_H + \sqrt{4 \beta^2 \rho_H^2 +
  (\Gamma-1)(\rho_H^2 - J^2)} \, ,
\end{equation*}
where $\beta \equiv (2 - \Gamma)/4$.
Next define
\begin{equation*}
  \chi_A \equiv \frac{(\Gamma - 1) J_A}{\Gamma p} \, , \qquad
  \chi^\varphi \equiv \frac{(\Gamma - 1) J^\varphi}{\Gamma p} \, , \qquad
  \chi^2 \equiv \chi_A \chi^A + \lambda^2 {\chi^\varphi}^2 \, .
\end{equation*}
We identify $\chi^A = W^2 v^A$ and $\chi^\varphi = W^2 v^\varphi$ and
hence with \eqref{eq:Lorentzrelation} we obtain
\begin{equation}
  \label{eq:Wfromcons}
  W^{-2} = \frac{1}{2 \chi^2} \left( \sqrt{1 + 4 \chi^2}
  - 1 \right) \, .
\end{equation}
This now enables us to calculate the velocities,
\begin{equation*}
  v_A = W^{-2} \chi_A \, , \qquad v^\varphi = W^{-2} \chi^\varphi \, .
\end{equation*}
The form of $W^{-2}$ in \eqref{eq:Wfromcons} guarantees that $|v_A|,
\, |v^\varphi| < 1$. This is most important since evolved speeds
greater than unity (i.e. greater than the speed of light) can easily
cause the numerical code to crash \cite{CN}.
  
Finally, we can calculate the specific enthalpy and  rest mass
energy density from \eqref{eq:prim2cons} and \eqref{eq:EOS_spec},
\begin{equation*}
  h = \frac{J^\varphi}{\sigma v^\varphi W} \, , \qquad 
  \rho = \frac{\Gamma p}{(\Gamma-1)h} \, .
\end{equation*}

%%% Local Variables: 
%%% mode: latex
%%% TeX-master: "master"
%%% End: 

\section*{Acknowledgments}

We thank Anita Barnes for several useful discussions.
OR thanks the Max Planck Institute for Gravitational Physics (Albert
Einstein Institute), Golm, Germany for his inclusion in their Visitor
Programme.
OR acknowledges gratefully financial support from the Engineering and
Physical Sciences Research Council, the Gates Cambridge Trust and
Trinity College Cambridge.  

\bibliography{refs}

\end{document}